\documentclass{scrartcl}

\usepackage{graphicx}
\usepackage{natbib}
\usepackage{amsmath}
\usepackage{subfigure}

\newsavebox{\astrutbox}
\sbox{\astrutbox}{\rule[-5pt]{0pt}{20pt}}

\title{Turbulence evolution in MHD plasmas}

\author{M. Wisniewski \thanks{Lehrstuhl f\"ur Astronomie, Universit\"at W\"urzburg, Emil-Fischer Str. 31, D-97074 W\"urzburg}, R. Kissmann \thanks{Institut f\"ur Astro- und Teilchenphysik, Universit\"at Innsbruck} and F. Spanier \thanks{Lehrstuhl f\"ur Astronomie, Universit\"at W\"urzburg, Emil-Fischer Str. 31, D-97074 W\"urzburg}
  \thanks{Email address for correspondence: fspanier@astro.uni-wuerzburg.de}}

\begin{document}

\maketitle

\begin{abstract}
  Turbulence in the interstellar medium has been an active field of
  research in the last decade. Numerical simulations are the tool of
  choice in most cases. But while there are a number of simulations on
  the market some questions have not been answered finally. In this
  paper we are going to examine the influence of compressible and
  incompressible driving on the evolution of turbulent spectra in a
  number of possible interstellar medium scenarios. We conclude that
  the driving not only has an influence on the ratio of compressible
  to incompressible component but also on the anisotropy of
  turbulence.
\end{abstract}

\section{Introduction}

Turbulence in the context of the interstellar medium (ISM) was for the
first time discussed by \citet{VonWeizsaecker1951}. This discussion,
however, was soon abandoned by the scientific community. Until the
late seventies there was not much progress in the field of
interstellar turbulence. Only when the observational techniques became
better at this time it became clear that there were power law
correlations between different structures in the ISM. At first
turbulence was only connected to the smallest scales of the ISM but
with improved  observations it was found that turbulence was present at all spatial scales. A
corresponding spectrum of the density fluctuations can, e.g., be found
in \citet{ArmstrongEtal1995}.

Besides the well known density fluctuation spectra also
velocity power spectra may be obtained. One possible method for this
has been presented by \citet{pogo} and was applied by
\citet{padoan}. The connection between the velocity and density
observations are not fully understood, as pointed out by
\citet{klessen}. For the case of the magnetic field structure
functions may be observed with radio telescopes
\citep[e.g.,][]{haverkorn}. For a more detailed account of the
observational evidence for interstellar turbulence see
\citet{ElmegreenScalo2004}.

Since there are yet no in-situ measurements of the
interstellar medium, a number of questions is still not
resolved. One of those is the question of
compressibility. Many theories assume an incompressible medium, yet
this is not compatible with the density fluctuation observations in
the ISM. In the wave picture of interstellar turbulence
\citep{spanier1, spanier2} it may well be assumed that up to certain wave numbers a major
compressible component is present. Another issue which is not yet
fully answered is the question of the driving of the turbulence. The energy balance of the
interstellar medium suggests that supernovae are the driving force
and energy is dissipated at small scales. Whether this driving is
compressible or incompressible is not clear. On the one hand the driving scenario involving supernova shock waves is connected to compression, since the shocks themselves are compressed. On the other hand
streaming cosmic rays may be also a source of driving especially through
incompressible Alfv\'en waves \citep{skilling}.

In the course of this paper we will come back to the main theories for
anisotropic MHD turbulence and review recent simulations. We then
present the results of our simulations and their implications for the
understanding of compressible MHD turbulence.

In this paper we will at first discuss the current knowledge about the theory of plasma turbulence, then we will p
resent our numerical methods. The results are shown for short time scales of turbulent driving, the evolving spectru
m during driving and the saturated spectrum.

\section{Theory}

\subsection{Turbulence theory}

\label{sec:turb_theo}

From the theoretical side the discussion about a turbulence theory
started out from the investigation of incompressible turbulence described by the Navier-Strokes
equation. The early discussions by Kolmogorov \citep[see, e.g.,][]{Kolmogorov1941} used the assumption of spatially
homogeneous turbulence. Later on, however, it was found that intermittency had to be taken into account leading to a slightly different form of the velocity spectrum. For different intermittency models and a general overview of the turbulence see \citet{Frisch1995}. For the turbulence in a magnetised fluid the model by \citet{SheLeveque1994} was extended to include compressibility and a magnetic field. \citet{politano95} have discussed the role intermittency in the solar wind, proposing sheet like structures \citep[also discussed in][]{grauer94}.\\
The next important step is the inclusion of a background magnetic field. In \citet{shebalin83} numerical studies have been conducted showing anisotropies of MHD plasmas due to the background field.
This aspect was addressed by 
\citet{GoldreichSridhar1994}, where the authors use a heuristic
wave-model to get some idea of the anisotropy of the turbulence
spectrum of magnetic turbulence with a background field. The effects
of a background magnetic field, however, are by no means fully
understood. This is reflected by the debate started in \citet{goldreich}, where a discussion about the correct description of parallel and perpendicular cascades started. 
Another aspect is the question when turbulence can be described by a wave-interaction picture and when non-linear processes dominated instead.

  To resolve this question a number of simulations have been conducted
  in the last 15 years. The first numerical test has been conducted by \citet{cho2000}, a prominent example are the simulations by
  \citet{marongoldreich}, who consider simulations of decaying
  incompressible turbulence. The incompressible plasma attracts far
  more attention by modellers for a number of reasons: A prominent one
  is that incompressible plasmas may be simulated with spectral
  methods providing a spectrum extending to high wave numbers. Apart
  from that basic turbulence theories are mainly developed for
  incompressible fluids.  But also \citet{Chandran} has done
  additional work on compressible turbulence in the $\beta\ll 1$ limit
  for the evolved spectra. Due to the very low $\beta$ limit sound
  waves are neglected in these simulations. Here, however, we will
  show the importance of sound waves in the context of compressible
  plasmas. A work which is more closely related to ours is the one of
  \citet{kuznetsov}, where the authors investigate compressible
  plasmas with high and low values of $\beta$. In his approach only
  weak turbulence is considered. As mentioned above it is not yet
  clear if a wave-picture is applicable, but in the weak turbulence
  limit the wave-picture is explicitly assumed. Additionally a theoretical description including major cooling and damping mechanisms for the interstellar medium has been presented by \citet{lithwick2001}. Here different $\beta$ regimes are discussed.

\subsection{Excitation of MHD waves with compressible or incompressible distortions}
\label{induction}
  In \citet{federrath1,federrath2} a discussion similar to our discussion
regarding the influence of compressible versus incompressible driving has been
started. But this discussion was limited to hydrodynamic simulations of the
interstellar medium. A more recent paper \citep{federrath3} added the magnetic
field and discussed the effect of turbulent magnetic field growth.  We will
extend this discussion in the case of magnetised fluids to the anisotropies
present in the driven turbulence. This discussion has been continued by
\citet{lemaster}, who did simulations of MHD plasmas, but limited themselves to
incompressible driving but with variations of the peak wave number of the
driver. In their work it was shown that approximately a factor of 128 between
the driving peak and the total resolution is sufficient for convergence.
\subsubsection{Compressible driving}
We inject compressible disturbances isotropically into our simulation
domain. These disturbances can distort the magnetic field and so excite waves.
The induction equation shows us which waves can be expected.  Without loss of
generality the preferred direction of the magnetic field can be chosen to be
the $x$-direction (defined by the unit vector $\vec{e}_x$). For the induction
equation we find that:
\begin{equation}
  \frac{\partial \vec{B}}{\partial t}= - \nabla \times \vec{E} = \nabla  \times (\vec{\delta v} \times B   \vec{e}_{x})
\end{equation}
Where we used $\vec{E}=-\vec{\delta v}\times\vec{B}$ for the electric field
which holds for ideal plasmas.  If the compressible fluctuations are
in $x$-direction no magnetic field will be injected as $(\delta v 
\vec{e_{x}} \times \vec{B} \cdot \vec{e_{x}})=0 $.  If the distortion
is perpendicular to the background field (e.g. $\vec{\delta v}= v
  \vec{e}_{y}$) this results in a distortion of the magnetic field in
$x$-direction:
\begin{eqnarray}
\frac{\partial \vec{B}}{\partial t}&=&i k \vec{e}_{y} \times (\delta v \vec{e}_{y} \times B\vec{e}_{x})\\\nonumber
\frac{\partial \vec{B}}{\partial t}&=& i k \vec{e}_{y}  \times (- v) B \vec{e}_{z}\\
\frac{\partial \vec{B}}{\partial t}&=& -i \vec{e}_{x} k  v  B
\label{FMW1}
\end{eqnarray}
Where we used $\nabla=i k \vec{e}_{y}$ as for compressible waves $\vec{k}$ is parallel to $\vec{\delta v}$

For the case $\vec{\delta v}= v \vec{e}_{z}$ we also find a distortion of the magnetic field in x-direction:
\begin{eqnarray}
\frac{\partial \vec{B}}{\partial t}&=& i \vec{e}_{x} k  v  B
\label{FMW2}
\end{eqnarray}
As the distortion of the magnetic field is perpendicular to $\vec{k}$, we would expect to find compressible waves that propagate in perpendicular direction with respect to the mean magnetic field. We can also identify these waves as for MHD-waves the propagation in the perpendicular direction is only possible for fast magnetosonic waves.

\subsubsection{Incompressible driving}
Here we want to figure out if Alfv\'en waves being the
only incompressible waves in MHD-plasmas can be generated form those
incompressible fluctuations. As for incompressible fluctuations
$\vec{k}$ is parallel to the background field $\nabla=k
\vec{e}_{x}$. So we find for the magnetic field:
\begin{equation}
  \frac{\partial \vec{B}}{\partial t}
  = k \vec{e}_{x}  \times (\delta v  \vec{ e}_{y} \times B  \vec{e}_{x}) = \vec{e}_{y} k v B
\label{Alf1}
\end{equation}
or
\begin{equation}
  \frac{\partial \vec{B}}{\partial t}
  =k \vec{e_{x}}  \times (\delta v \vec{ e}_{z} \times B \vec{e}_{x}) = \vec{e}_{z} k v B
  \label{Alf2}
\end{equation}
As the distortion of the magnetic field is perpendicular to $\vec{k}$
this results in Alfv\'en waves propagating parallel with respect to
the mean magnetic field.

\subsection{Three wave interaction}
\label{chap:3www}
Three wave interaction is used to describe the interaction of
compressible and incompressible waves in a weakly turbulent plasma
and has been described in the first place by
  \citet{chinwentzel}.  Applications to high- and low-beta plasmas are
  discussed in \citet{vs05}.\\For our investigations where we either
drive with pure compressible or pure incompressible fluctuations one
has to put the question with which process compressible fluctuations
are generated from incompressible fluctuations and the other way
around. As this happens for small turbulent amplitudes we claim that
three wave interactions are the key process.\\Here a validation is
carried out how Alfv\'en waves are generated by fast magnetosonic
waves. It is also shown what can be predicted about the propagation
direction of the resulting Alfv\'en waves. The calculation of the
generation of fast magnetosonic waves from Alfv\'en waves can be done
analog.
\\
In principle Alfv\'en waves can be generated from fast
magnetosonic waves by one of these two processes:
\begin{eqnarray}
  A^{+}+F^{-}&\leftarrow&A^{\pm}\label{3w1}\\
  A^{+}+A^{-} &\rightarrow& F^{\pm}\label{3w2}
\end{eqnarray}
Here we will restrict the discussion
  to the special case of perpendicular propagating fast magnetosonic
waves with $\beta \ll 1$. For this we find the phase velocity of
these waves to be:
\begin{eqnarray}
v_{ph}\approx (\sqrt{1+\beta})v_{A}
\label{FMW_phase}
\end{eqnarray}
In this case the
  reaction equation Eq. \ref{FMW_phase} provides no valid solution
  when computing the resonance condition (i.e., the energy equation).
For reaction equation \ref{3w2} , however, we find the
following resonance condition:
\begin{align}
  v_{A}(k_{A,\parallel}^{+}+k_{A,\parallel}^{-})&=
  v_{A}(\sqrt{1+\beta})k_{F}^{\pm}\\
  v_{A}(k_{A,\parallel}^{+}+k_{A,\parallel}^{-})&=
  v_{A}(\sqrt{1+\beta})((k_{A,\parallel}^{+}-k_{A,\parallel}^{-})^{2}+k_{F,\perp}^{\pm})^{1/2}\label{3w4}
\end{align}
with
\begin{align}
  k_{A,\parallel}^{+}-k_{A,\parallel}^{-}
  =\pm k_{F,\parallel}^{\mp}\label{3w3}
  k_{A,\perp}^{+}+k_{A,\perp}^{-}
  =\pm k_{F,\perp}^{\mp}
\end{align}
As the fast magnetosonic waves propagate in the perpendicular
direction $k_{F,\parallel}=0$ and so with equation
\ref{3w4} and \ref{3w3}
  we find for the parallel component of the fast magnetosonic waves:
\begin{eqnarray}
k_{A,\parallel}^{+}=k_{A,\parallel}^{-}=\frac{1}{2}k_{F,\parallel}\sqrt{1+\beta}
\end{eqnarray}

For the perpendicular component of the fast magnetosonic waves we find
\begin{eqnarray}
\langle k_{A,\perp}^{\pm}\rangle=\frac{1}{2}k_{F}
\end{eqnarray}
With that we find that
\begin{eqnarray}
\tan{\theta}=\frac{k_{A,\parallel}}{k_{A,\perp}}=\frac{\frac{1}{2}k_{F}\sqrt{1+\beta}}{\frac{1}{2}k_{F}}=\sqrt{1+\beta}
\end{eqnarray}
where $\theta$ is the angle between the propagation direction
of the fast magnetosonic waves and the Alfv\'en waves.  As $\beta \ll
1$ it can be followed that $\tan{\theta}\approx 1$. This gives us a
preferred propagation direction for the Alfv\'en waves close to
$\theta=45$ degree.

A similar study has been undertaken by \citet{chandranprl} for the low-$\beta$ case taking into account also fast magnetosonic wave. \citet{lithwick2001} have discussed systems with high and low $\beta$, their approach is an analytical one.

\section{Numerical Simulations}

\subsection{MHD equations}
For our studies we used the ideal MHD equations in a periodic domain, where the system is
closed by an isothermal equation of state. Since our analysis is a
principle one we will give all variables in non-dimensional,
normalised units. For that purpose all variables are split into a
dimensionless variable of order unity and a normalisation
constant. Here we use four independent variables for the
normalisation: The length of the simulation domain $L$, the mass of the
hydrogen atom $m_0$, a typical number density $n_0$, and the temperature of the
system $T_0$ which directly relates to the speed of sound $c_s$.
Therefore length scales are normalised as $x = L \tilde x$, velocity as $u = u_0 \tilde u = c_s\tilde u$, density as $\rho = m_0 n_0\tilde\rho$ and so forth.
The resulting set of normalised MHD-equations is the following:
\begin{eqnarray}
  \frac{\partial \rho}{\partial t}  &=& -\nabla \vec s  \\
  \frac{\partial \vec s}{\partial t}  &=&-\nabla \left( \frac{\vec s \vec s }{\rho} +    \left( p + \frac{B^2}{2} \right)1-\vec B \vec B \right)+\vec F  \\
  \frac{\partial \vec B}{\partial t}&=&  -\nabla \left(\frac{\vec s \vec B -\vec B \vec s}{\rho} \right) \\
  p &=& \frac{c_s ^2}{u_0 ^2}\rho \stackrel{(here)}{=} \rho
\label{iso}
\end{eqnarray}
Here $\rho$ is the mass density, $\vec s= \rho \vec v$ is the momentum
density and $\vec{B}$ indicates the magnetic induction and $p$ is the
thermal pressure. The isothermal equation of state -- Eq. (\ref{iso})
-- is simplified due to our choice for the normalisation constants.

In the following we will classify our results on the basis of $\beta$ we used for our simulations.
We only use it for the initialisation of the unperturbed background magnetic fields. It is defined as the fraction of the thermal and magnetic pressure.

\subsection{Numerical model}
For the time evolution of the MHD-equations we use a second order
CWENO (Centrally Weighted Essentially Non-Oscillatory) algorithm (see
e.g. \citep{2000math......2133K}). It is very convenient for the
simulation of shocks as we find them in fluids and plasmas. In the
mathematical description these shocks are so called ``Riemann
Problems'' and their accurate numerical treatment with a Godunov
solver results in high numerical costs. The CWENO scheme avoids to
solve these Riemann Problems directly by averaging over the
appropriate fractions of every cell. Additionally oscillations are
suppressed by the reconstruction. This results in a fast scheme for
the treatment of the MHD equations with passable small dissipation.
This CWENO method is combined with a third order Runge-Kutta algorithm
which has been chosen because of the small memory costs with
comparatively high resolution in time, due to the fact that only one
additional field has to be buffered.

Besides the direct numerical simulation (DNS) technique also
  the large eddy simulation. This is for example demonstrated in
  \citet{cherny07,cherny09} for the case of decaying turbulence. We
  are refraining from using this method since it requires a number of
  implications on the turbulent cascade on small scales which we are
  not sure of.

\subsection{Turbulence driver}
\label{turb_driver}
For the external stirring of the fluctuations we define a
function $f$ in Fourier space with
\begin{eqnarray}
f_k=s\cdot k^{-7/4}\exp(2\pi \imath p)
\end{eqnarray}
where ``$s$'' and ``$p$'' are random numbers between zero and one. The
random number generator for the former of these obeys a Gaussian
distribution, whereas the latter is uniformly distributed. This
function $f_k$ yields the Fourier space distribution of the velocity
fields used for driving the fluctuations in our simulations. After
transforming to configuration space $f_k \longrightarrow f_x$ we make
the distinction between the cases of compressible and incompressible
driving. For the former we use $\vec{\delta v} =\nabla f_x$, whereas
for the latter we employ $\vec{\delta v} =\nabla \times \vec f$.
As $\vec f$ has, thus, to be a vectorial function in the case of
incompressible driving we define each component separately with
different random numbers.  Each component of the wave-vector $\vec k$
in our Fourier space spectra ranges from one to eight yielding a
maximum absolute value of $\mid\vec k\mid\le  13$.  The perturbations are set up
in a form such that no net momentum is added to the numerical
domain. The energy input is normalised to an estimate for the
average energy input by supernovae into the ISM of our Galaxy into a
volume of the size of the simulation domain. The resulting driving spectrum is isotropic and it yields $v^2(k)\propto k^{-3/2}$ which yields a production spectrum harder than the anticipated inertial spectrum.

\label{en_spec}

Finally we define the compressible and incompressible
energy spectra. Both are calculated in
Fourier space on the basis of the velocity fields with
\begin{align}
  P_{comp}(k)&\equiv\frac{1}{2}\left\| \widehat{k}\cdot \vec{v}(\vec k)  \right\|^2 \\
  P_{incomp}(k)&\equiv\frac{1}{2}\left\|  \widehat{k}\times \vec{v}(\vec k)  \right\|^2.
\end{align}
where $\widehat{k}$ is the normalised wave vector. The overall energy spectrum of the velocity
fields is
\begin{eqnarray}
  P_{K}(k)\equiv \frac{1}{2}v^2(k)
\end{eqnarray}
where the sum of $P_{comp}(k)$ and $P_{incomp}(k)$ is $P_{K}(k)$.
For the following discussion we have to make an important
  differentiation: $P_{K}$ is the power spectrum of velocity fields in the whole system, whereas $P_{K}^{*}$ is the power spectrum of the velocity fields of the driver. This is important because the turbulent energy of the velocity fields of the whole system doesn't have to accord to the driving spectrum. This will be important in chapter \ref{pk_shorttime}. Additionally we define the omnidirectional energy spectrum of the magnetic fields with
\begin{eqnarray}
  P_{B}\equiv\frac{1}{2\bar\rho}\delta B^2(k).
\end{eqnarray}

\subsection{Simulation setup}
For all our investigations we use in principle the same simulation
setup: We start out from an unperturbed plasma with a homogeneous
background magnetic field and inject with each time step either
compressible or incompressible energy with our turbulence driver
explained in section \ref{turb_driver}. We do this for compressible
and also for incompressible driving and both runs are performed for
$\beta$=0.01 and $\beta$=10. This enables us to analyse the influence
of the different drivings on as well the highly magnetised as the
quite hydrodynamic plasmas. So the basis of the following research are
those four runs. With the simulation output we gain during the
simulation we calculate the turbulent energy spectra defined in
section \ref{en_spec}.

\section{Results}
Here we will discuss three distinct topics relating to our
  simulation results: First we investigate the physics on short and
intermediate timescales where the fluctuations of the velocity fields
excite modes. This is done in section
\ref{shorty}. Then the time evolution of the energies
is studied in section \ref{Energieverlauf}. Finally an investigation
of the convergence range is done in section \ref{chap:converge}. In order to 
quantify the anisotropy we use a scalar parameter which is derived
from 2D spectra. The parameter is described in detail in \ref{app:aniso}.

\subsection{Short time scales}
\label{shorty}
This investigation is of special interest because it can be observed
how the first waves are generated for compressible as well as for
incompressible driving and how these waves interact with each
other. On the basis of this research we can determine which processes dominate the turbulence
 for small amplitudes of the fluctuations. When taking the supernova injection scenario for
  granted, the build-up of turbulence is a transient phenomenon which may even be observed
  close in time and space to the energy source. One observable could
  be accelerated particles in the upstream of
  supernova and their respective electromagnetic emission.

For the investigation of the two dimensional spectra we need to keep
in mind that after this short amount of time we have to distinguish
between the driving region that is dominated by the driving spectrum
injected for each time step and the part of the spectrum that can only
be reached via a cascade.  After this short amount of time the
cascading process has only taken place to some extent so we
concentrate mostly on the driving region.  Via the intersection points
of the contour lines with the $k_{\perp}$- and $k_{\parallel}$-axes we
can access the anisotropy of the turbulent fields.

\subsubsection{$P_B$}
\label{PB}
\begin{figure}
\centering
\setlength{\unitlength}{0.00038\textwidth}
\subfigure[$\beta$=0.01 incompressible driving]{
  \begin{picture}(1000,803)(-100,-50)
    \put(-60,380){\rotatebox{90}{$\ln k_{\perp}$}}
    \put(330,-40){$\ln k_{\parallel}$}
    \includegraphics[width=900\unitlength]{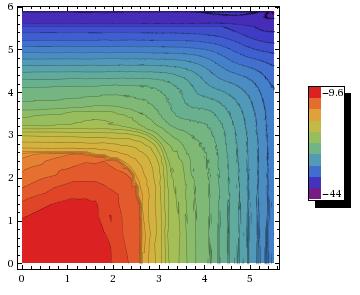}
  \end{picture}
  \label{fig:subfig2.1}
}
\subfigure[$\beta$=0.01 compressible driving]{
  \begin{picture}(1000,803)(-100,-50)
    \put(-60,380){\rotatebox{90}{$\ln k_{\perp}$}}
    \put(330,-40){$\ln k_{\parallel}$}
    \includegraphics[width=900\unitlength]{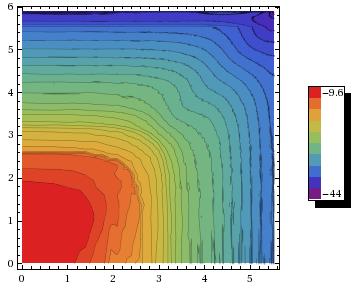}
  \end{picture}
  \label{fig:subfig2.2}
}
\subfigure[$\beta$=10 incompressible driving]{
  \begin{picture}(1000,803)(-100,-50)
    \put(-60,380){\rotatebox{90}{$\ln k_{\perp}$}}
    \put(330,-40){$\ln k_{\parallel}$}
    \includegraphics[width=900\unitlength]{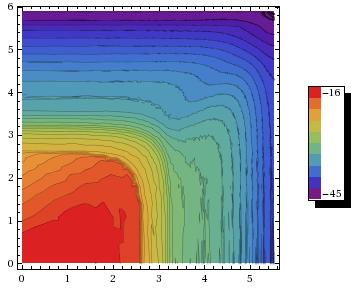}
  \end{picture}
  \label{fig:subfig2.3}
}
\subfigure[$\beta$=10 compressible driving]{
  \begin{picture}(1000,803)(-100,-50)
    \put(-60,380){\rotatebox{90}{$\ln k_{\perp}$}}
    \put(330,-40){$\ln k_{\parallel}$}
    \includegraphics[width=900\unitlength]{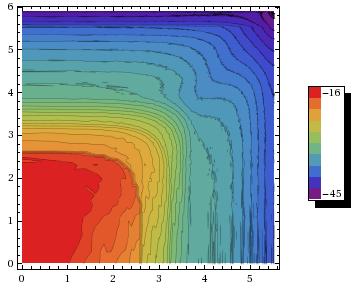}
  \end{picture}
  \label{fig:subfig2.4}
}
\caption{Turbulence spectra $P_{B}$ after $t=5\cdot10^{-3}$ simulated on a $512^{3}$ grid. The color denote the natural logarithm of the numerical energy.}
\end{figure}
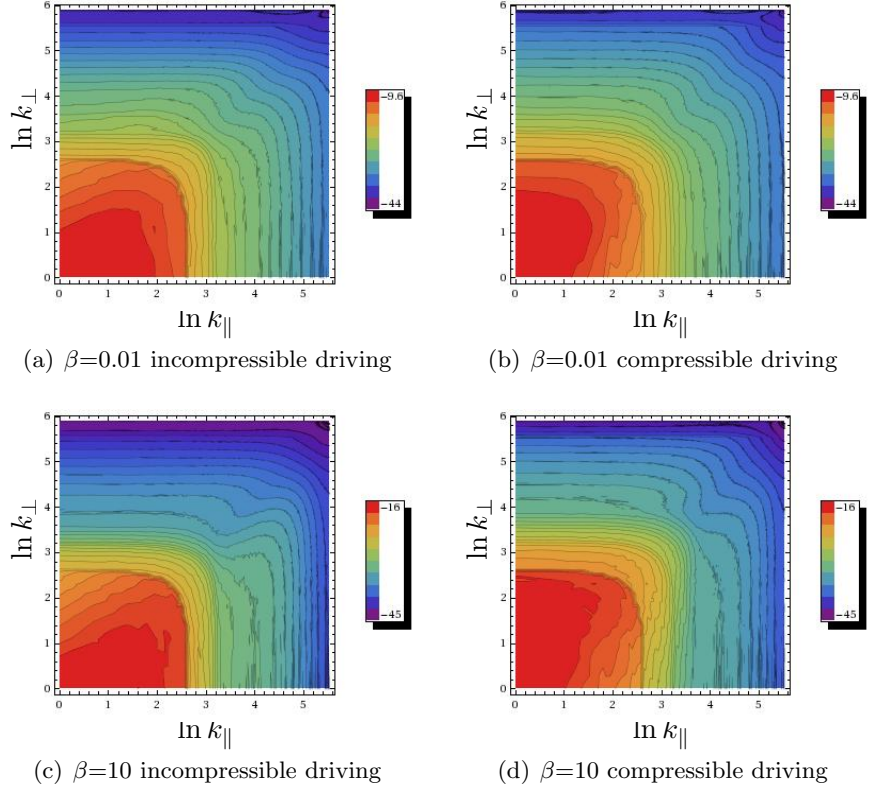

Figures \ref{fig:subfig2.1} to \ref{fig:subfig2.4} show the two
dimensional spectra of $P_{B}$.  We see for both drivings and values of the
plasma-$\beta$ a very anisotropic spectrum ($k_\perp^\text{aniso}(1) =
2$) within the driving range.  The spectra show a remarkable feature:
For incompressible driving contour lines in the driving range rise from the $k_\perp$ axis 
(following the diagonal), while for compressible driving the contour lines rise
from the $k_\parallel$ axis. The slope of the contour lines indicates a spectrum which
has in parts a positive exponent with respect to $k$.

There is an explanation for this behaviour, which may be deduced from the fact, that
the difference exist between compressible and incompressible fluctuations. For the case
of incompressible fluctuations, the vector $\vec{\delta v}$ is perpendicular to the 
wave vector $\vec{k}$, while for compressible fluctuations, the both vectors are 
aligned. If we examine now the induction equation for incompressible fluctuations (Eq. 
\ref{Alf1}) moving in parallel direction, we find
\begin{equation}
\frac{\partial \vec{B}}{\partial t}  = \nabla\times(\delta v \vec{e}_y \times B_0 \vec{e}_x)
\end{equation}
this yields for the turbulent velocity spectrum
\begin{equation}
\frac{\partial \vec{B}}{\partial t} = k_\parallel k^{-3/4} B_0
\end{equation}
This yields a rising spectrum in parallel spectrum as seen above. It also explains the crossing
of the contour line and the $k_\perp$ axis. For the compressible driving the
explanation follows the same line, but is essentially different, since the vector 
$\vec{\delta v}$ has a different orientation.

We can make further distinction between $\beta=0.01$ and $\beta=10$: If we examine
the 2D spectra closely we find a cutoff in some of the spectra, which may be identified
by very close contour lines (as for example in Fig. \ref{fig:subfig2.1}). For compressible driving 
we find a strong cutoff for $\beta=10$, while it seems to have dissolved already for
$\beta=0.01$. Since the induction equation makes no distinction between the two values 
of $\beta$ it seems that another physical process is taking place here. By examining
$P_{Comp}$ and $P_{Shear}$ we will show later that this may be caused by a mixture
of  Alfv\'en and fast magnetosonic waves.

\begin {figure}
\begin{center}
\setlength{\unitlength}{0.00038\textwidth}
\subfigure[$\beta$=0.01 incompressible driving]{
\begin{picture}(1000,803)(-100,-50)
    \put(330,-40){$x$}
    \put(-40,330){\rotatebox{90}{$y$}}
    \includegraphics[width=900\unitlength]{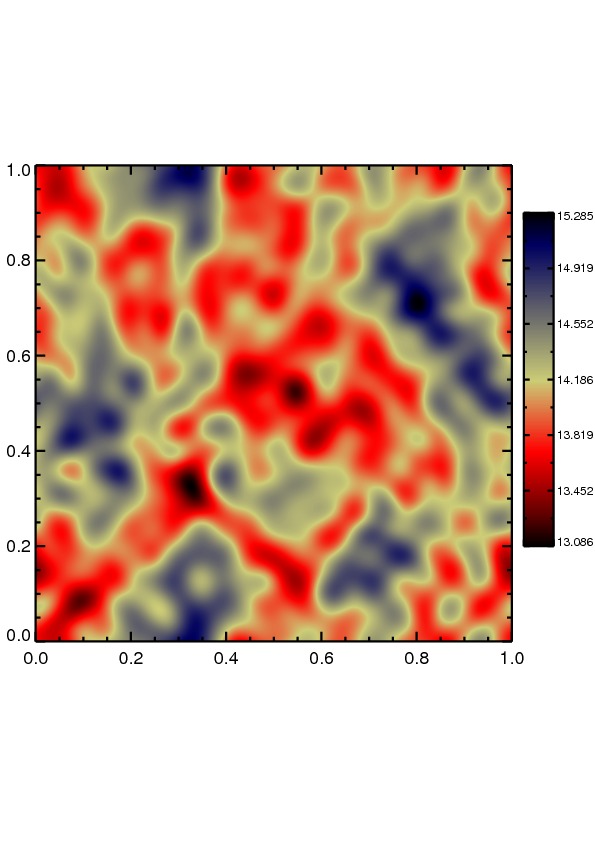}
 \end{picture}
  \label{fig:subfigFelder_PB_t50b001C1}
}
\subfigure[$\beta$=0.01 compressible driving]{
\begin{picture}(1000,803)(-100,-50)
    \put(330,-40){$x$}
    \put(-40,330){\rotatebox{90}{$y$}}
    \includegraphics[width=900\unitlength]{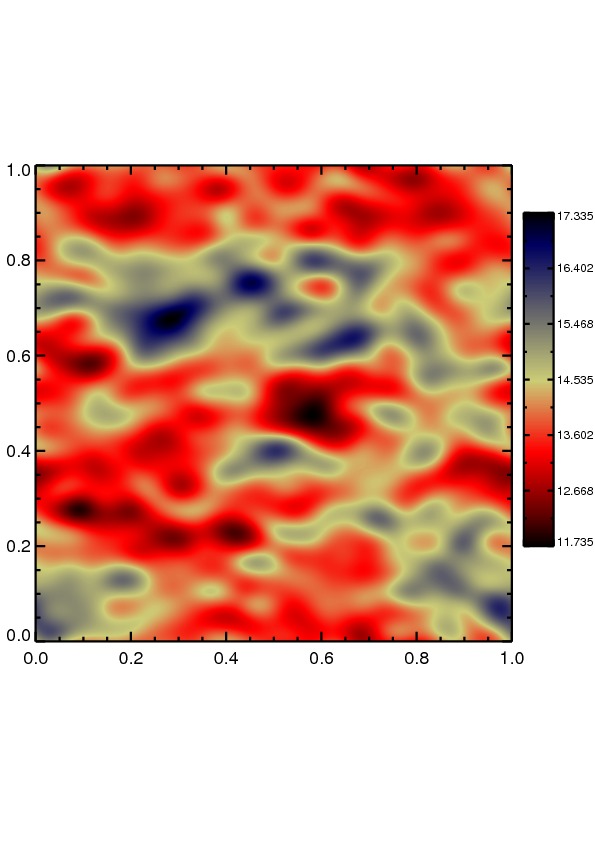}
 \end{picture}
  \label{fig:subfigFelder_PB_t50b001C2}
}
\subfigure[$\beta$=10 incompressible driving]{
\begin{picture}(1000,803)(-100,-50)
    \put(330,-40){$x$}
    \put(-40,330){\rotatebox{90}{$y$}}
    \includegraphics[width=900\unitlength]{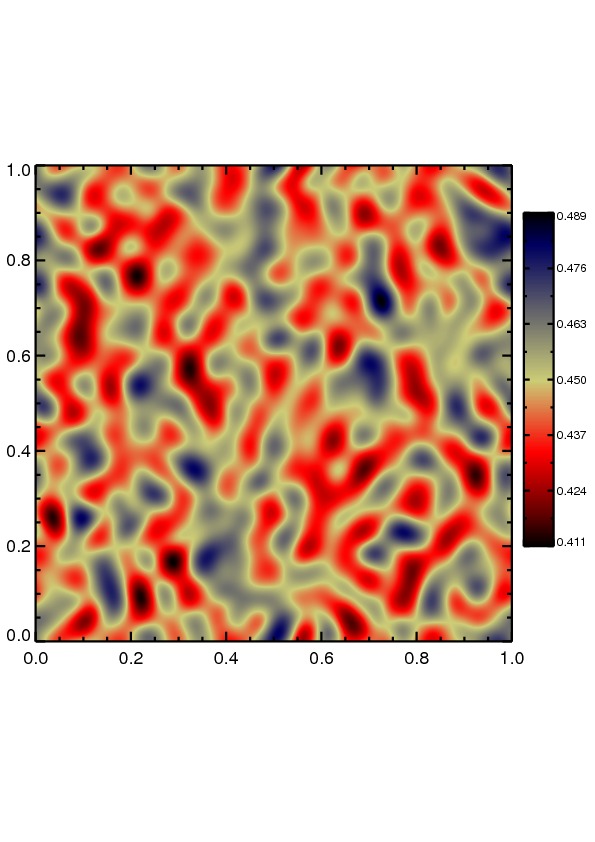}
  \end{picture}
  \label{fig:subfigFelder_PB_t5010C1}
}
\subfigure[$\beta$=10 compressible driving]{
\begin{picture}(1000,803)(-100,-50)
    \put(330,-40){$x$}
    \put(-40,330){\rotatebox{90}{$y$}}
    \includegraphics[width=900 \unitlength]{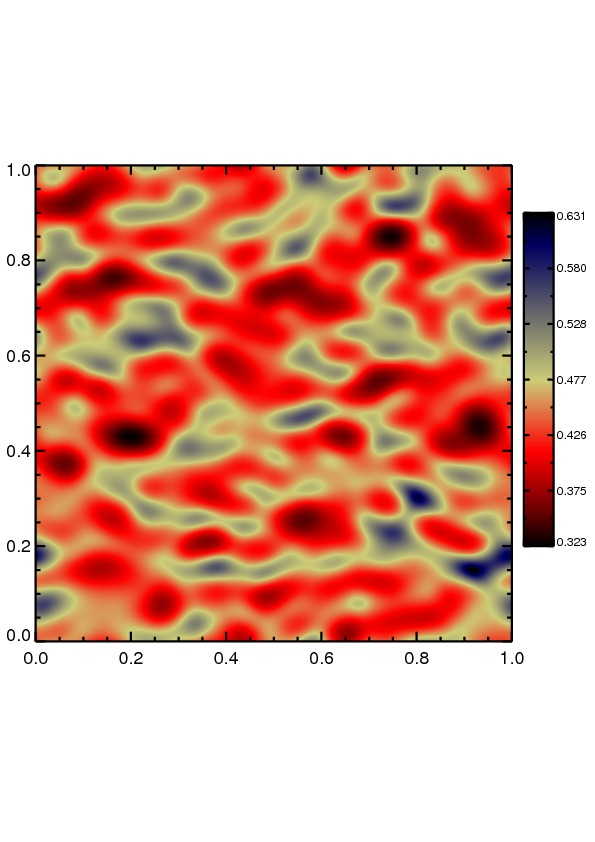}
  \end{picture}
  \label{fig:subfigFelder_PB_t50b10C2}
}
\end{center}
\caption{Turbulent magnetic fields after $t=5\cdot 10^{-3}$ in real
  space on a $512^{3}$ grid.}
\end {figure}

In \ref{fig:subfigFelder_PB_t50b001C1} to \ref{fig:subfigFelder_PB_t50b10C2} the
magnetic field after a normalised time of $5\cdot 10^{-3}$ can
be seen. One may now compare the eddy size and spectrum (as done for
example in \citealt{beresnyak2009}) to see at least qualitatively that the
elongation of the eddies is correlated with the cutoff seen in the
two-dimensional spectra. This is of course not unexpected.

\subsubsection{$P_K$}
\label{pk_shorttime}
\begin{figure}
\centering
\setlength{\unitlength}{0.00038\textwidth}
\subfigure[$\beta$=0.01 incompressible driving]{
  \begin{picture}(1000,803)(-100,-50)
    \put(-60,380){\rotatebox{90}{$\ln k_{\perp}$}}
    \put(330,-40){$\ln k_{\parallel}$}
    \includegraphics[width=900\unitlength]{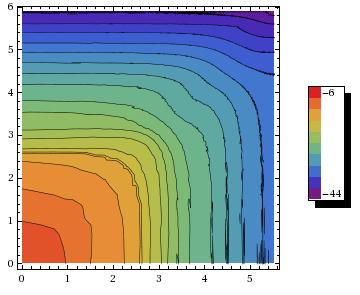}
  \end{picture}
  \label{fig:subfig2.1_a}
}
\subfigure[$\beta$=0.01 compressible driving]{
  \begin{picture}(1000,803)(-100,-50)
    \put(-60,380){\rotatebox{90}{$\ln k_{\perp}$}}
    \put(330,-40){$\ln k_{\parallel}$}
    \includegraphics[width=900\unitlength]{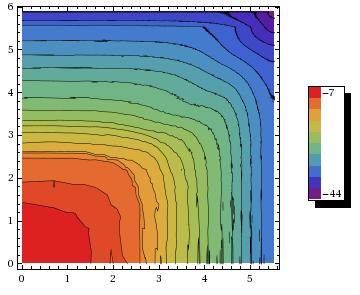}
  \end{picture}
  \label{fig:subfig2.2_a}
}
\subfigure[$\beta$=10 incompressible driving]{
  \begin{picture}(1000,803)(-100,-50)
    \put(-60,380){\rotatebox{90}{$\ln k_{\perp}$}}
    \put(330,-40){$\ln k_{\parallel}$}
    \includegraphics[width=900\unitlength]{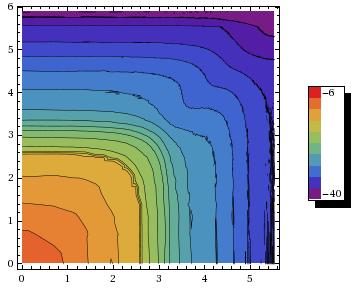}
  \end{picture}
  \label{fig:subfig2.3_a}
}
\subfigure[$\beta$=10 compressible driving]{
  \begin{picture}(1000,803)(-100,-50)
    \put(-60,380){\rotatebox{90}{$\ln k_{\perp}$}}
    \put(330,-40){$\ln k_{\parallel}$}
    \includegraphics[width=900\unitlength]{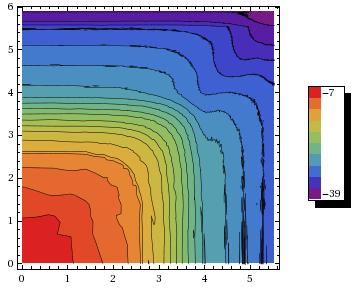}
  \end{picture}
  \label{fig:subfig2.4_a}
}
\caption{Turbulence spectra $P_{K}$ after $t=5\cdot 10^{3}$ simulated on a $512^{3}$ grid.}
\end{figure}
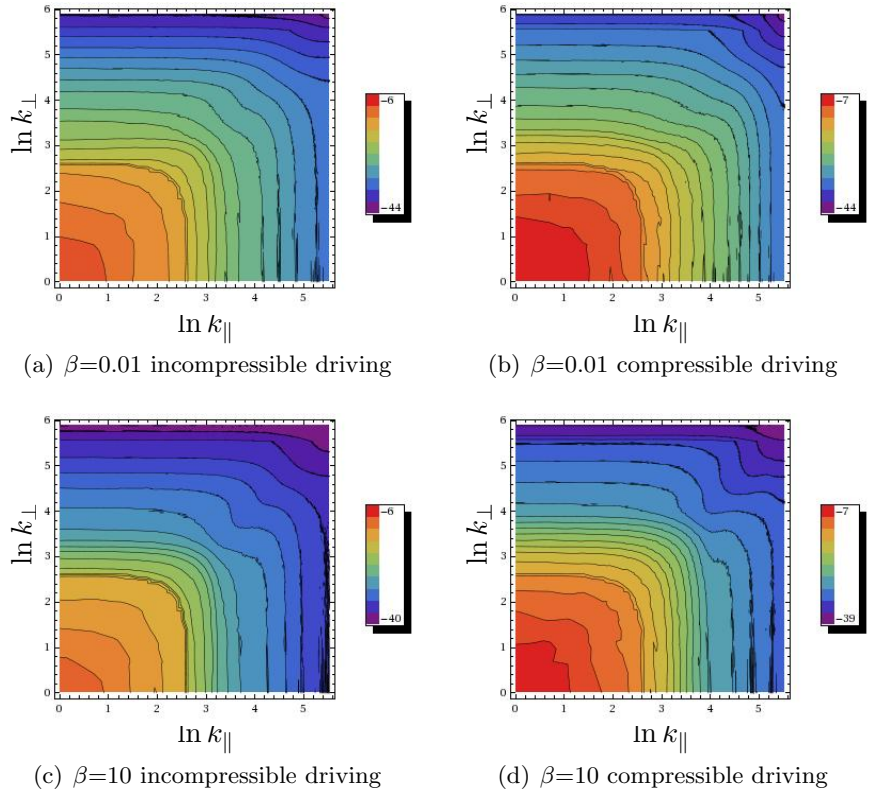

Figure \ref{fig:subfig2.1_a} to \ref{fig:subfig2.4_a} show the two
dimensional spectra of $P_{K}$ for $\beta$=0.01 and $\beta$=10 for
both compressible and incompressible
driving.\\For $\beta$=0.01 and incompressible driving we see in the
driving region of  $P_{K}$ a perpendicular preferred direction ($k_\perp^\text{aniso}(1.5)=1.8$) whereas for
$\beta$=0.01 and compressible driving we find a parallel preferred
direction ($k_\perp^\text{aniso}(1.5)=1.5$) although $P_{K}^{*}$ is for both cases isotropic. The term ``preferred direction'' is used here when the two dimensional spectra are anisotropic. The contour lines of a spectrum with a parallel preferred direction cross the axes for higher values of $k_{\parallel}$ than $k_{\perp}$. For the contour lines of a spectrum with a perpendicular preferred direction it's the other way around.
For $\beta$=10 we see for compressible as well as for
incompressible driving a very homogeneous driving spectrum. This can
be understood quite easily. We drive the turbulence with compressible
or incompressible fluctuations in the velocity fields. As shown in
chapter \ref{PB} this results in an excitation of MHD-waves in the
driving range, so turbulent energy of the velocity fields is
transferred into magnetic energy. In the direction of preferred
excitation of waves this energy is missing in the velocity fields and
so in $P_{K}$. This is why for the case of incompressible driving and
$\beta=0.01$ we find Alfv\'en waves propagating parallel to the mean
magnetic field and so a perpendicular preferred direction in
$P_{K}$. For compressible driving we excite fast magnetosonic waves
perpendicular to the mean magnetic field which results in a parallel
preferred direction of $P_{K}$.\\For $\beta=10$ although the same is
in all likelihood happening this effect can't be seen, because the
magnetic fields are much weaker and so less waves are excited and
through this less energy of $P_{K}$ is transformed into
$P_{B}$. Hereby the missing energy in $P_{K}$ in one of the preferred
directions can't be detected in $P_{K}$.

\subsubsection{$P_{Shear}$ and $P_{Comp}$}

\begin{figure}
\centering
\setlength{\unitlength}{0.00038\textwidth}
\subfigure[$\beta$=0.01 incompressible driving]{
  \begin{picture}(1000,803)(-100,-50)
    \put(-60,380){\rotatebox{90}{$\ln k_{\perp}$}}
    \put(330,-40){$\ln k_{\parallel}$}
    \includegraphics[width=900\unitlength]{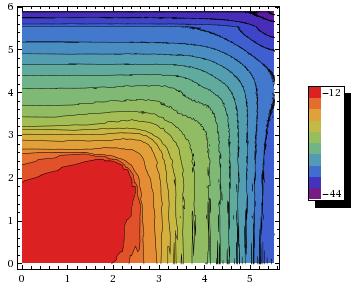}
  \end{picture}
  \label{fig:subfig5.1}
}
\subfigure[$\beta$=0.01 compressible driving]{
  \begin{picture}(1000,803)(-100,-50)
    \put(-60,380){\rotatebox{90}{$\ln k_{\perp}$}}
    \put(330,-40){$\ln k_{\parallel}$}
    \includegraphics[width=900\unitlength]{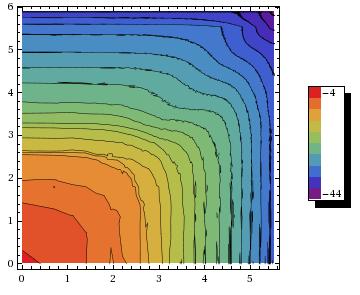}
  \end{picture}
  \label{fig:subfig5.2}
}
\subfigure[$\beta$=10 incompressible driving]{
  \begin{picture}(1000,803)(-100,-50)
    \put(-60,380){\rotatebox{90}{$\ln k_{\perp}$}}
    \put(330,-40){$\ln k_{\parallel}$}
    \includegraphics[width=900\unitlength]{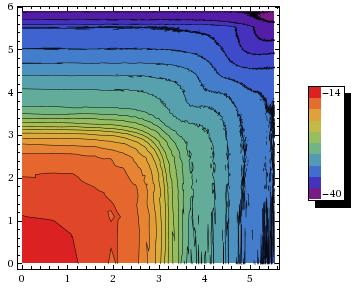}
  \end{picture}
  \label{fig:subfig5.3}
}
\subfigure[$\beta$=10 compressible driving]{
  \begin{picture}(1000,803)(-100,-50)
    \put(-60,380){\rotatebox{90}{$\ln k_{\perp}$}}
    \put(330,-40){$\ln k_{\parallel}$}
    \includegraphics[width=900\unitlength]{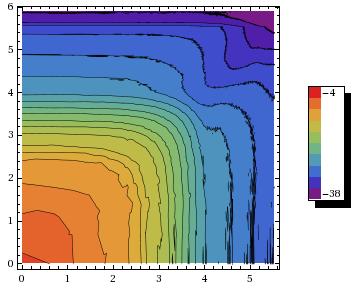}
  \end{picture}
  \label{fig:subfig5.4}
}
\caption{$P_{Comp}$ after $t=5\cdot 10^{-3}$ simulated on a $512^{3}$ grid.}
\end{figure}
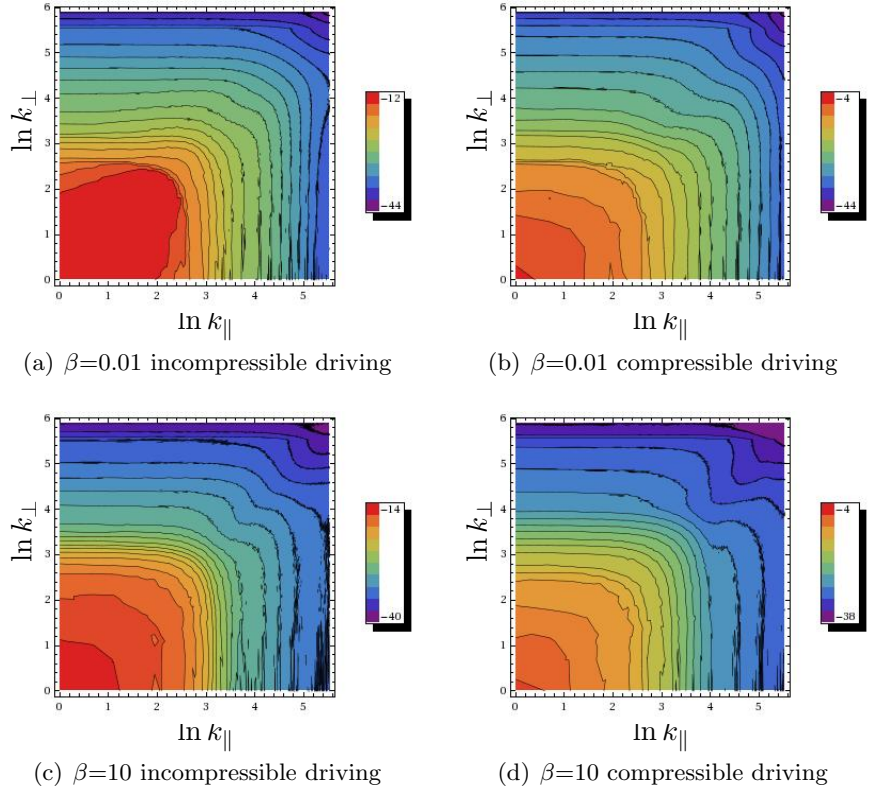

\begin{figure}
\centering
\setlength{\unitlength}{0.00038\textwidth}
\subfigure[$\beta$=0.01 incompressible driving]{
  \begin{picture}(1000,803)(-100,-50)
    \put(-60,380){\rotatebox{90}{$\ln k_{\perp}$}}
    \put(330,-40){$\ln k_{\parallel}$}
    \includegraphics[width=900\unitlength]{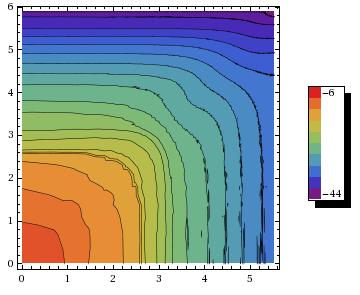}
  \end{picture}
  \label{fig:subfig4.1}
}
\subfigure[$\beta$=0.01 compressible driving]{
  \begin{picture}(1000,803)(-100,-50)
    \put(-60,380){\rotatebox{90}{$\ln k_{\perp}$}}
    \put(330,-40){$\ln k_{\parallel}$}
    \includegraphics[width=900\unitlength]{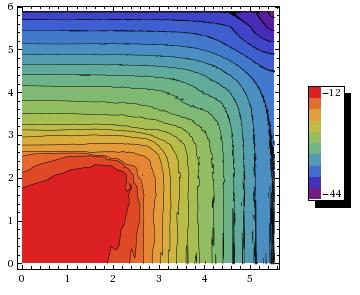}
  \end{picture}
  \label{fig:subfig4.2}
}
\subfigure[$\beta$=10 incompressible driving]{
  \begin{picture}(1000,803)(-100,-50)
    \put(-60,380){\rotatebox{90}{$\ln k_{\perp}$}}
    \put(330,-40){$\ln k_{\parallel}$}
    \includegraphics[width=900\unitlength]{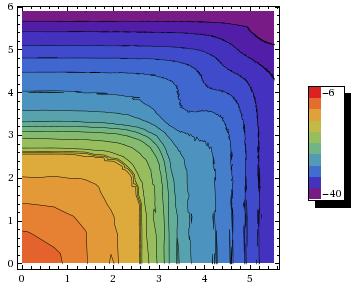}
  \end{picture}
  \label{fig:subfig4.3}
}
\subfigure[$\beta$=10 compressible driving]{
  \begin{picture}(1000,803)(-100,-50)
    \put(-60,380){\rotatebox{90}{$\ln k_{\perp}$}}
    \put(330,-40){$\ln k_{\parallel}$}
    \includegraphics[width=900\unitlength]{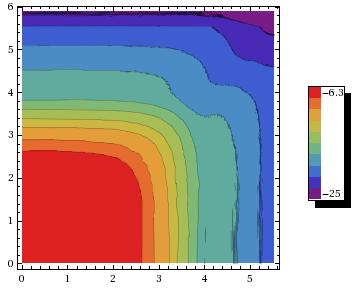}
  \end{picture}
  \label{fig:subfig4.4}
}
\caption{$P_{Shear}$ after $t=5\cdot 10^{-3}$ simulated on a $512^{3}$ grid.}
\end{figure}

Now we will investigate the two dimensional spectra of $P_{Comp}$ and
$P_{Shear}$. Figure \ref{fig:subfig5.1} to \ref{fig:subfig4.4} show
the two dimensional spectra of $P_{Comp}$ and $P_{Shear}$ at a
normalised time of $5\cdot 10^{-3}$. At this very early time the type
of driving dominates the turbulent energy of the velocity fields.  So
for $\beta=0.01$ the driving component shows of cause the same
behaviour as $P_{K}$ for $\beta=0.01$. We also find for compressible
driving a parallel preferred direction in $P_{Comp}$ ($k_\perp^\text{aniso}(2.2)=2.0$) as in this case
fast magnetosonic waves are generated propagation perpendicular to the
mean magnetic field and this energy comes from the turbulent energy of
the velocity components. For $P_{Shear}$ and incompressible driving we
find a perpendicular preferred direction ($k_\perp^\text{aniso}(1.4)=1.8$), as we have an excitation of
parallel propagating Alfv\'en waves. The non driven components -
$P_{Comp}$ for incompressible driving and $P_{Shear}$ for compressible
driving - show in principle the same diagonal symmetry. In chapter
\ref{chap:3www} it was shown that Alv\'en waves can be generated via
three wave interaction from fast magnetosonic waves propagating
perpendicular to the mean magnetic field and that these Alv\'en waves
propagate with an angel of 45 degree to the fast magnetosonic
waves. This prediction corresponds to the two dimensional spectrum of
$P_{Shear}$ for compressible driving. The same can be done for
$P_{Comp}$ and incompressible driving.\\For $\beta=10$ we find the
same $P_{Comp}$ and $P_{Shear}$ limitations as for $P_{K}$: Because of
the weak magnetic fields only few waves are generated and this effect
becomes as little noticeable as in the driven component as in
$P_{K}$. As the non-driven component is also absolutely isotropic, no
clear conclusion about the dominating physical process can be made.

  One should note that there should exist a connection between
  the $P_{comp}$ results and the density fluctuations through the
  continuity equation. Density fluctuations have been studied in
  detail by \citet{kowal07b}, where only solenoidal driving is used.

\subsection{Evolving spectrum}
\label{Energieverlauf}
\begin {figure}
\begin{center}
\subfigure[turbulent magnetic energy]{
  \includegraphics[width=0.45\textwidth]{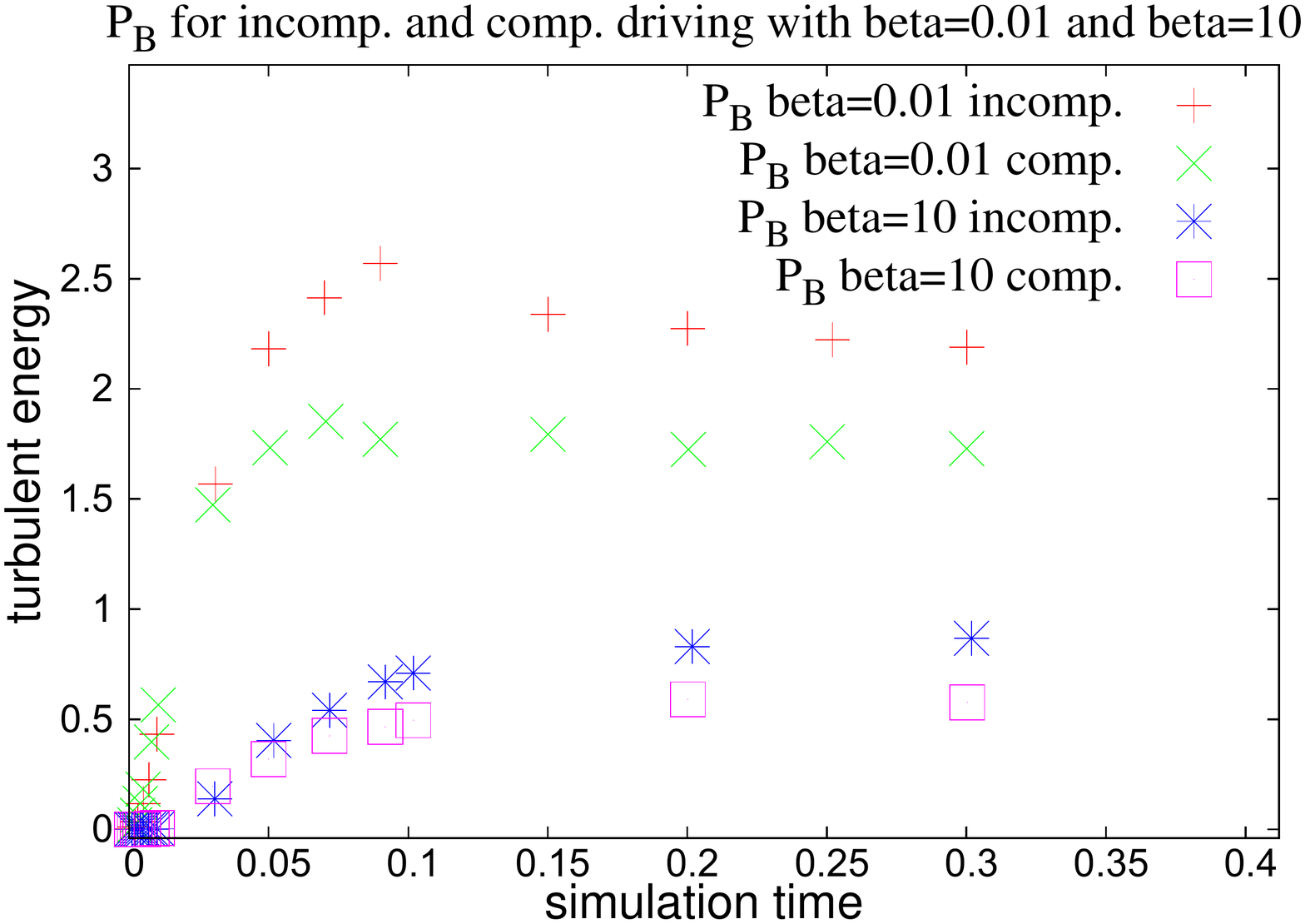}
  \label{fig:subfig0.01}
}
\subfigure[turbulent energy of the velocity fields]{
  \includegraphics[width=0.45\textwidth]{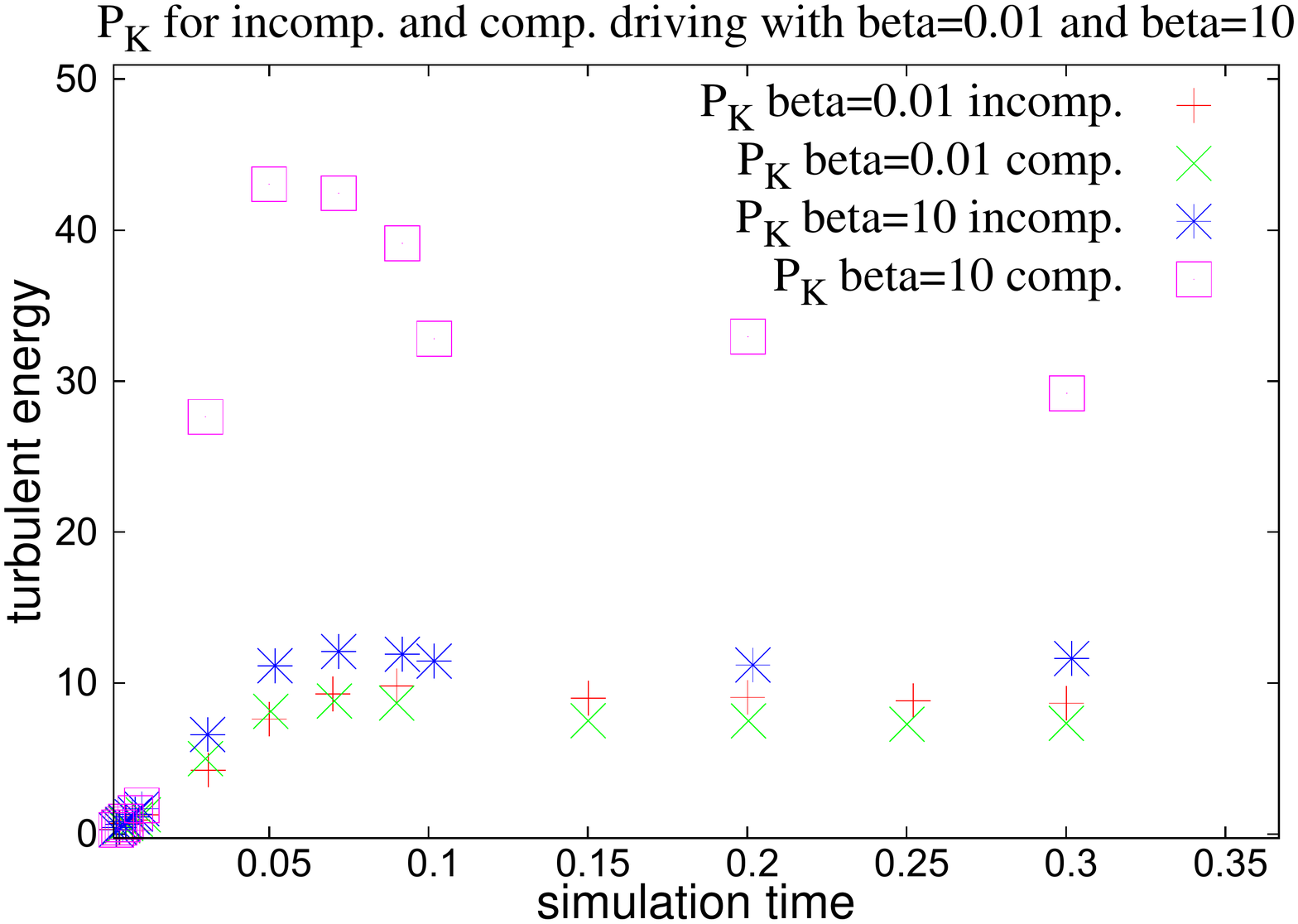}
  \label{fig:subfig0.02}
}
\end{center}
\caption{Turbulence spectra $P_{K}$ and $P_{B}$ as a function of the simulation time }
\end {figure}

In the following chapter $P_{K}$,$P_{Comp}$,$P_{Shear}$ and $P_{B}$
for the four different setups are plotted against the normalised time
(see figure \ref{fig:subfig0.01} and \ref{fig:subfig0.02} as well as
figure \ref{fig:subfig0.03} and \ref{fig:subfig0.04} ). It can be seen
that there is no turbulent energy in the fields at the beginning of
the simulation. As the turbulence is driven continuously through the
whole simulation more and more turbulent kinetic energy is injected in
the fields with time until at last
the turbulence saturates (this is when the input
  energy from the driving is balanced by the numerical
  dissipation). A detailed analysis about the
saturation range is given in section
\ref{chap:converge}. Here we are mainly concerned about the physics
before convergence is reached. The absolute amplitude of the energies
in the convergence range is influenced by numerical effects and is so
neglected in the discussion.

\subsubsection{Turbulent kinetic energy $P_{K}$}
\begin {figure}
\begin{center}
\subfigure[magnetic field lines for $\beta=10$ with compressible driving after t=0.05]{
  \includegraphics[scale=0.13]{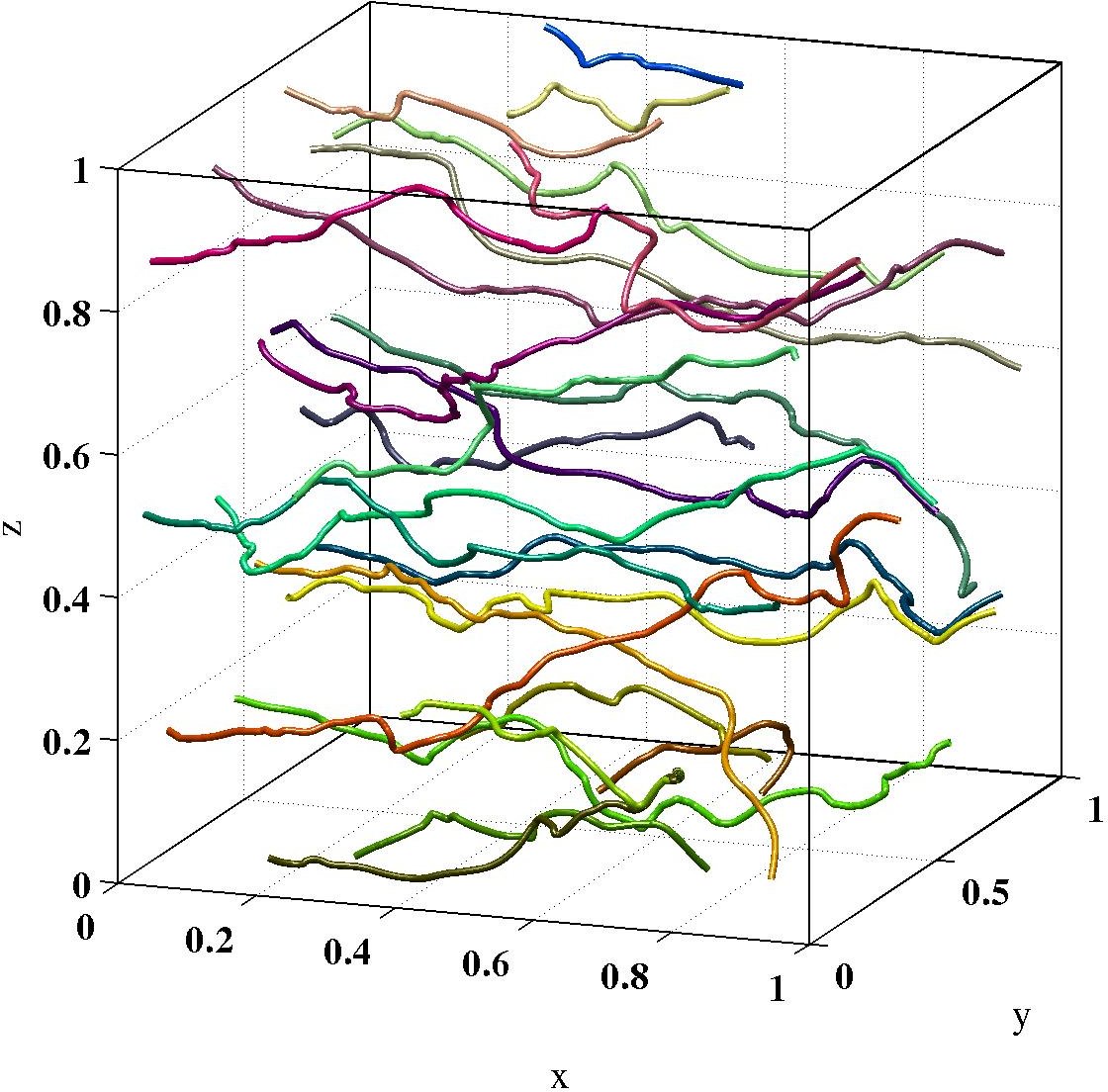}
  \label{fig:subfigFeldliniencompbeta10t500ts}
}
\subfigure[magnetic field lines for $\beta=10$ with compressible driving after t=0.3]{
  \includegraphics[scale=0.13]{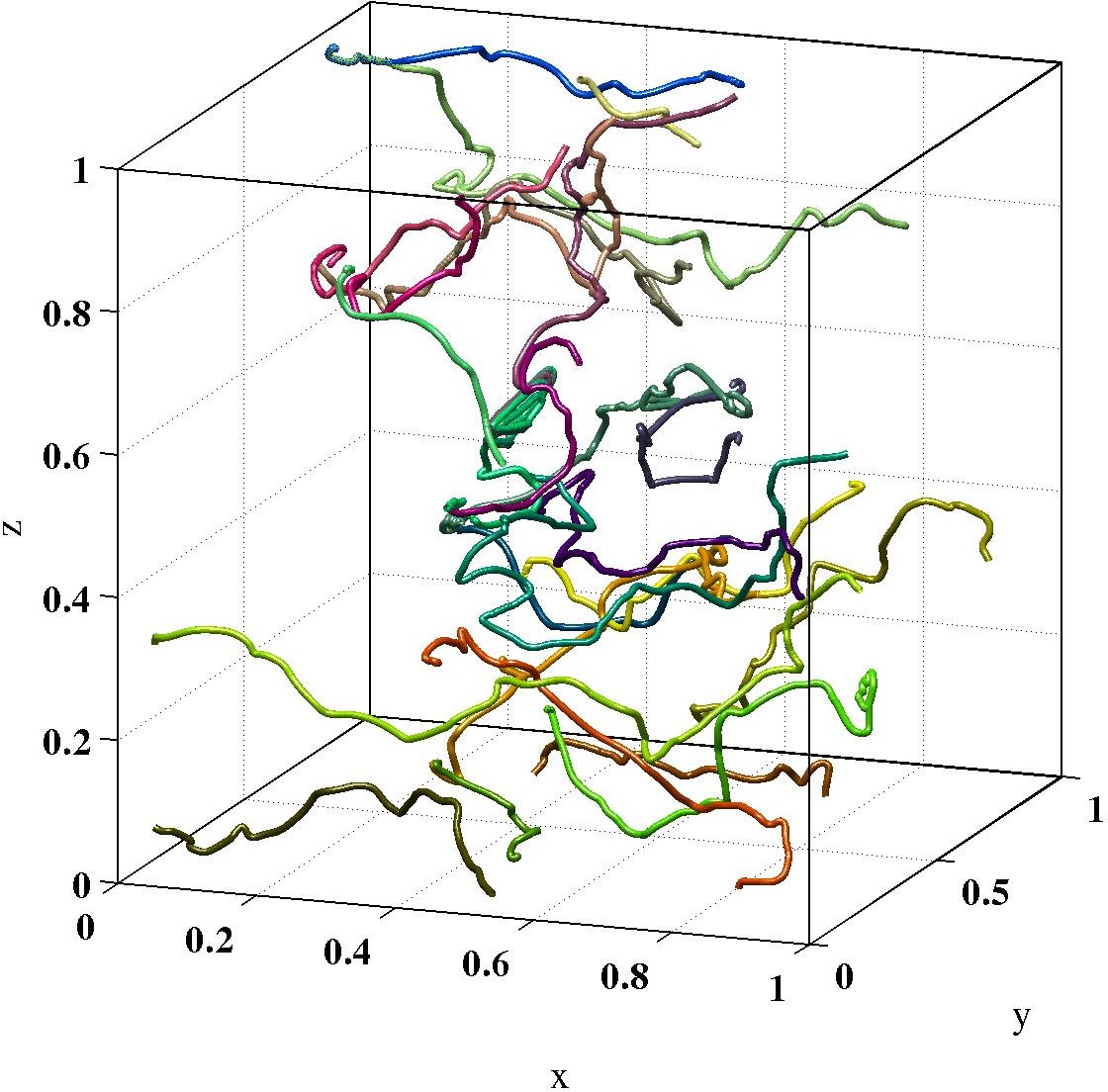}
  \label{fig:subfigFeldliniencompbeta10t3000ts}
}
\subfigure[magnetic field lines for $\beta=10$ with incompressible driving after t=0.05]{
  \includegraphics[scale=0.13]{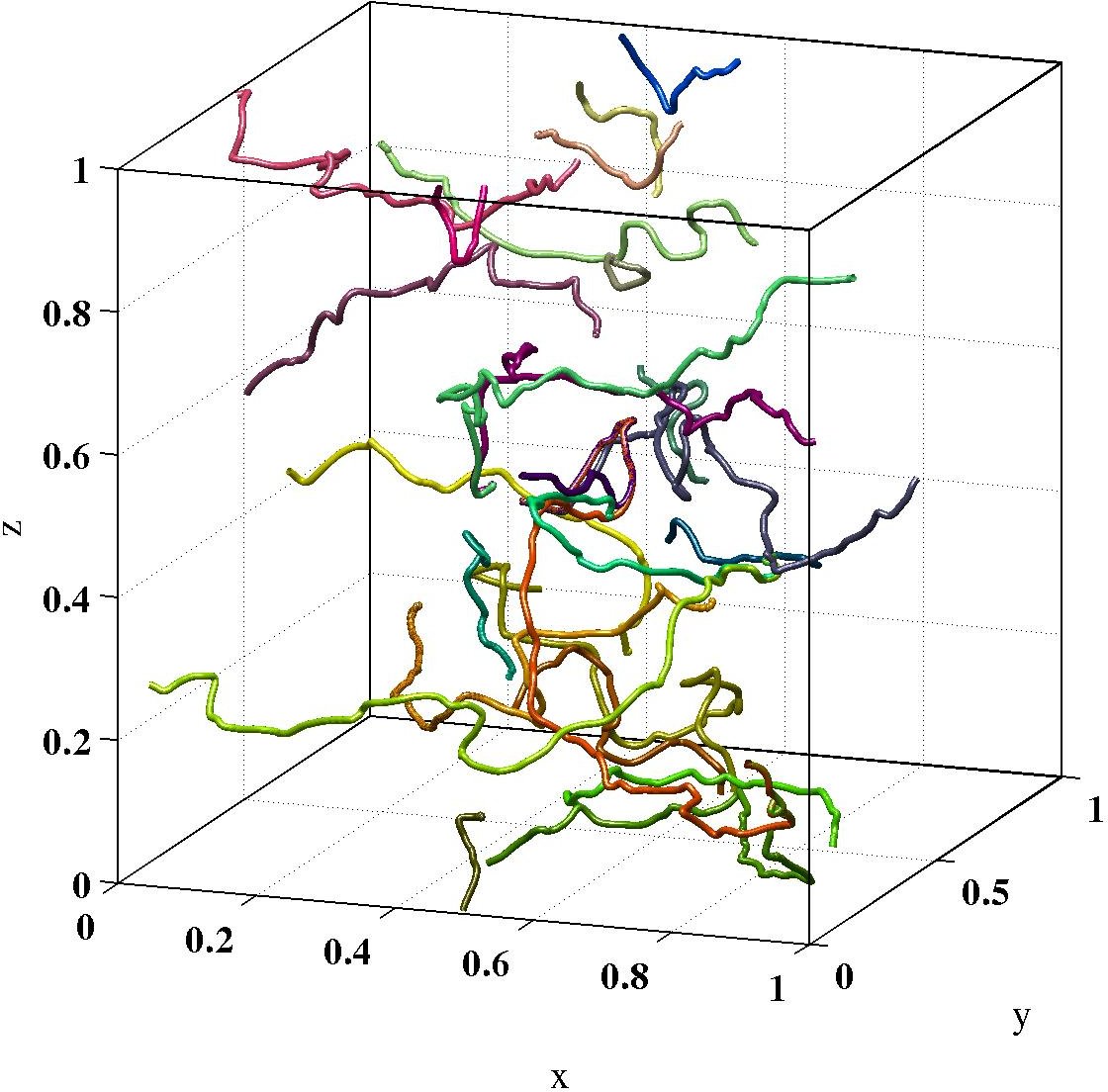}
  \label{fig:subfigFeldlinienincompbeta10t500ts}
}
\subfigure[magnetic field lines for $\beta=10$ with incompressible driving after t=0.3]{
  \includegraphics[scale=0.13]{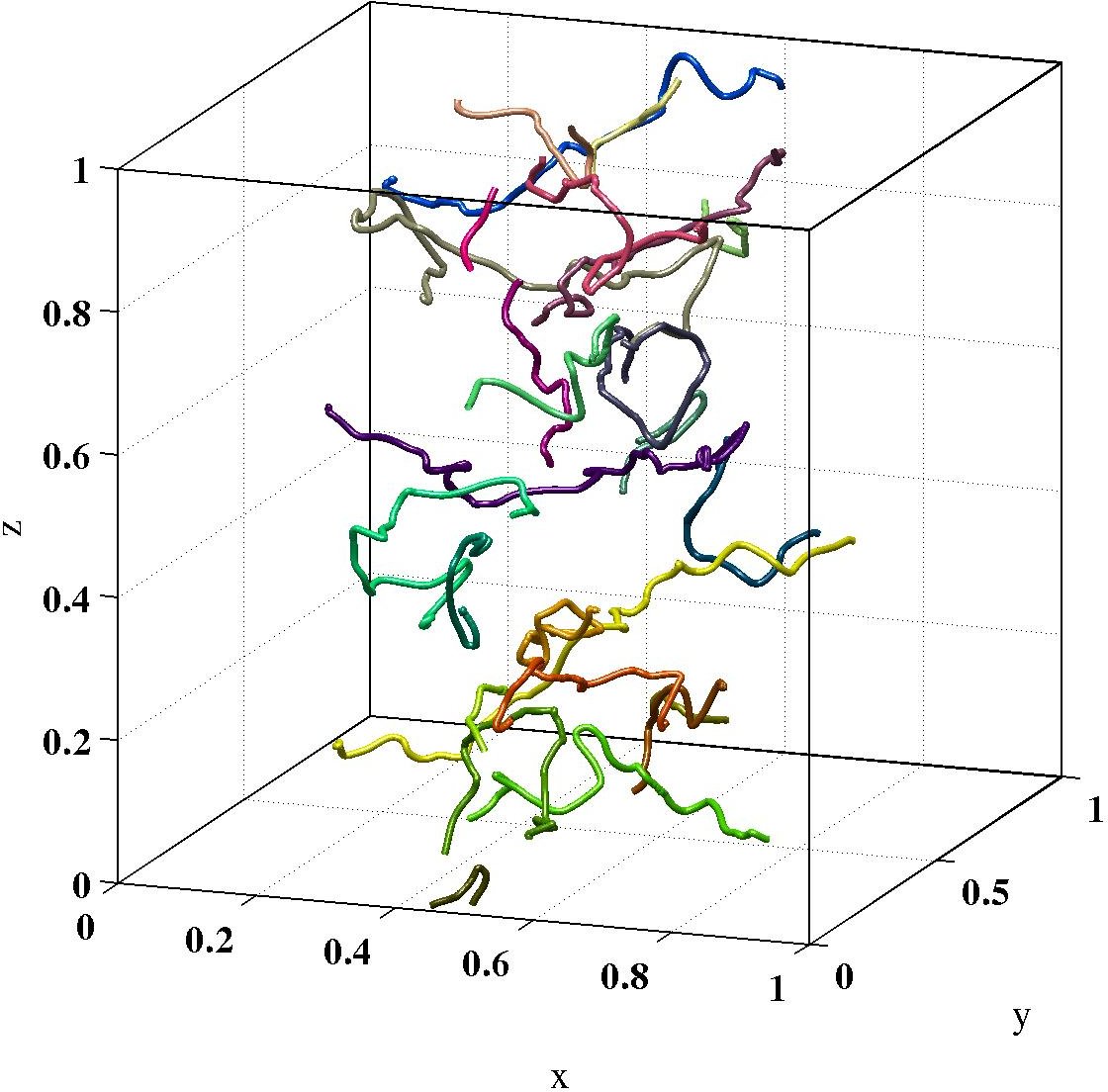}
  \label{fig:subfigFeldlinienincompbeta10t3000ts}
}
\end{center}
\caption{Magnetic field lines for $\beta=10$ with compressible and incompressible driving }
\end {figure}

The temporal evolution of $P_{K}$ differs fundamentally between
compressible and incompressible driving (see figure
\ref{fig:subfig0.02}). This is especially apparent for $\beta=10$. For
compressible driving we find a distinct maximum for $P_{K}$ at
t$\simeq$ 0.05. After that the energy decreases before finally
saturation is reached.
\\
What causes the formation of this maximum? In principle we find for
both - compressible and incompressible fluctuations - the same behaviour: If
the magnetic field lines are bent, the cascade and so the dissipation
of the energy in the system is more efficient. From the data we can
see that for compressible driving perpendicular fluctuations are generated
that decay into the parallel direction which leads to an isotropic
spectrum. To do this they need a component parallel to the background
magnetic field. If the background field is quite tangled
the fluctuations have a parallel component locally and can decay.  For
incompressible driving parallel fluctuations are generated that decay into
perpendicular direction and so they need a perpendicular component to
the background field. This is why they as well can only decay in a
tangled 
field. The formation of the maximum can occur if the driving energy
will bend the magnetic field lines only insufficiently. So for quite a
long time there won't be a cascade and the resulting dissipation of
the energy. Hereby lots of energy is accumulated until finally the
field lines are bent enough to build up the cascade. Then the
accumulated energy for small $\vec{k}$ can at last be reduced by the
cascade and the combined dissipation. If the driving energy bends the
magnetic field lines efficiently from the beginning the cascade will
build up from the beginning. Through this no energy is accumulated and
no maximum will occur.\\The magnetic field lines for compressible and
incompressible driving at this maximum (t=0.05) and after convergence
is reached are shown in figure
\ref{fig:subfigFeldliniencompbeta10t500ts} to
\ref{fig:subfigFeldlinienincompbeta10t3000ts}. It can be seen that for
the case of compressible driving the magnetic field lines are still
quite smooth at the maximum whereas in the convergence range they are
quite tangled. For the case of incompressible driving
however the field lines are already quite tangled at
t=0.05. This can be understood, when considering how the driving acts on the fluid. For a incompressible
  driving we are basically adding eddy-like motions to our fluid. This
  means that we get some shear-flows, which will certainly bend the
  background field lines. When adding a compressible velocity field,
  however, the background field even remains undisturbed for some of
  the input wave-modes. This shows, that an incompressible driving
  field will tangle the field lines much more efficiently. Thus, we
  find for $\beta=10$ that with the incompressible driving field the
  field is tangled so early that such an energy storage as in the case of the
  compressible driving can not form at all.

This tangling of the magnetic field lines can also be accessed
quantitatively. For this we computed the spatial averages of the
quantity
\begin{equation}
  a = 1 - 2 \mu^2
\end{equation}
where $\mu$ is the angle between the local and the initial magnetic
field. For this Parameter we find alignment along the initial magnetic
field direction for $a = -1$, where a magnetic field that is
perpendicular to the initial one would yield $a=1$. Due to the fact
that there is a higher degree of freedom for the perpendicular
direction an isotropic distribution would yield $a = 1/3$. With this
in mind we can discuss the results for our simulations.

\begin{figure}
  \setlength{\unitlength}{0.00045\textwidth}
  \begin{picture}(1000,961)(-100,-100)
    \put(440,-65){$\theta$}
    \put(-75,390){\rotatebox{90}{$f(\theta)$}}
    \includegraphics[width=800\unitlength]{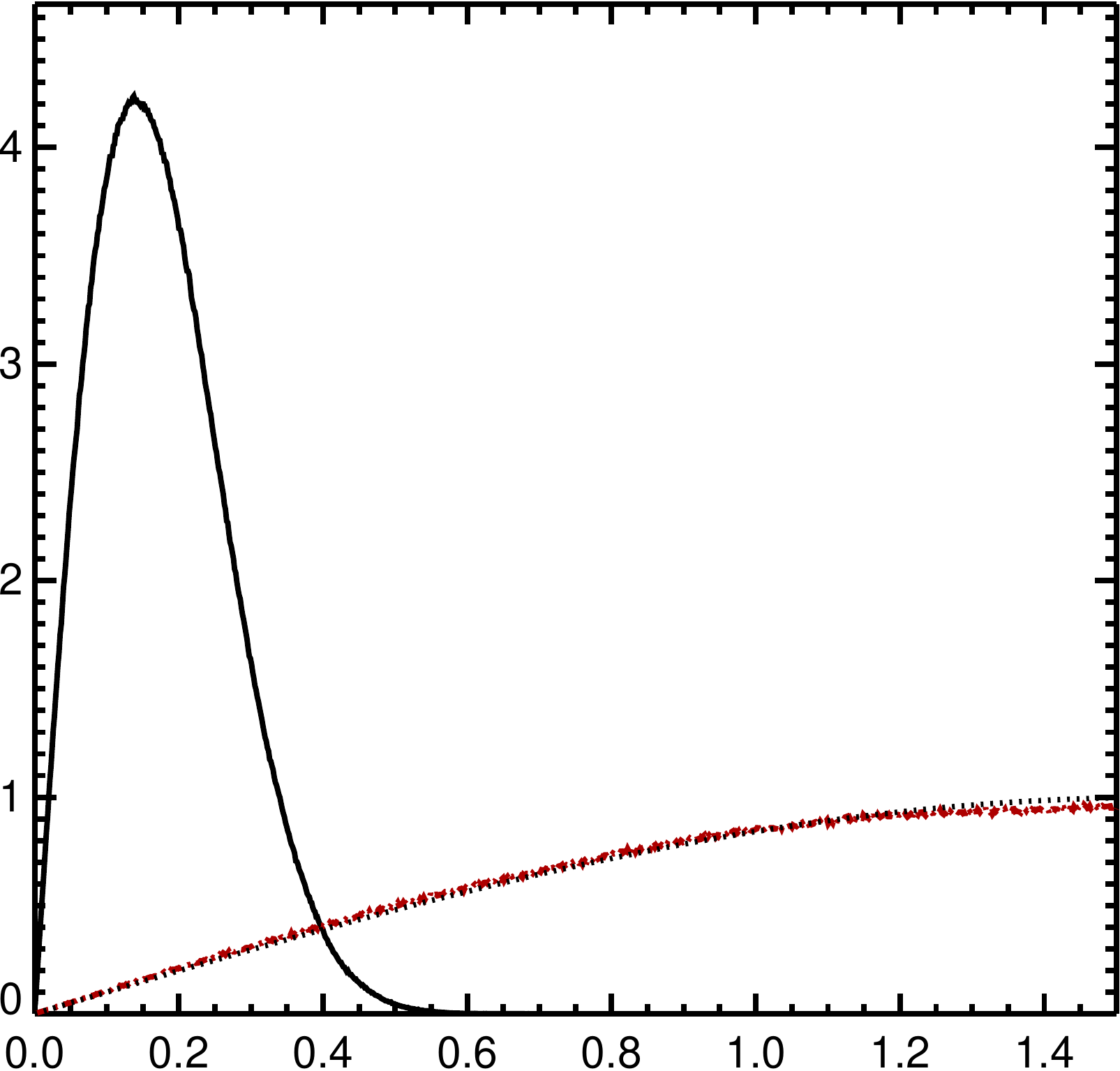}
  \end{picture}
  \hfill
  \begin{picture}(1000,961)(-100,-100)
    \put(440,-65){$\theta$}
    \put(-75,390){\rotatebox{90}{$f(\theta)$}}
    \includegraphics[width=800\unitlength]{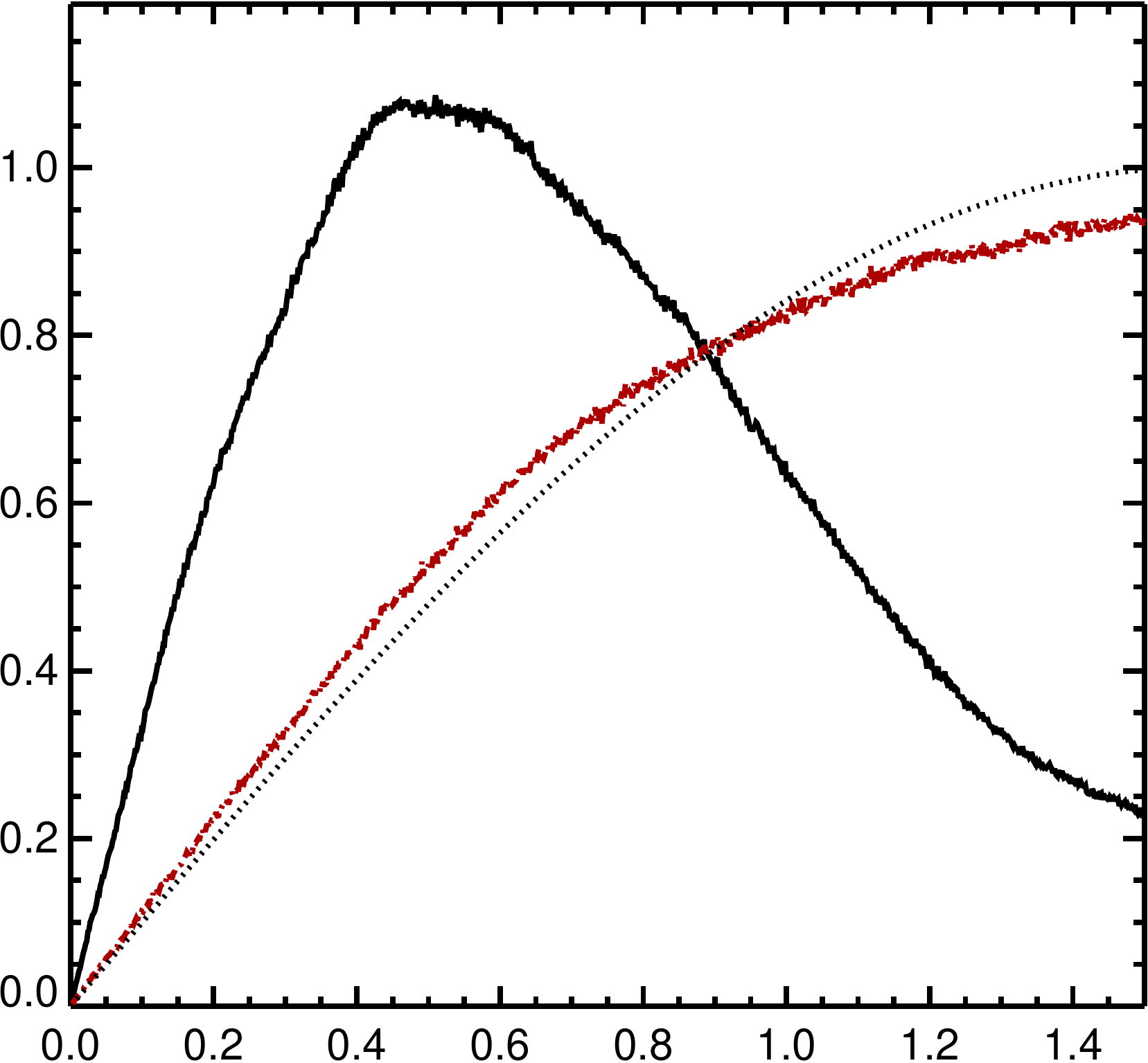}
  \end{picture}
  \caption{\label{FigThetaDist} Distribution function of the magnetic
    field angle for different times and different numerical simulation
    runs -- all results are shown for $\beta = 10$. On the left we
    show results for the incompressibly driven turbulence for $t=0.01$
    (black curve) and for $t=0.3$ (red curve). On the right the
    results for compressible (black curve) and incompressible driving
    (red curve) are compared for $t=0.05$. In both plots the dashed
    line indicates homogeneous distribution.}
\end{figure}

For an initial plasma-$\beta$ of $\beta = 0.01$ the
alignment-parameter $a$ even for the saturated turbulence at $t=0.3$
never exceeds values of -0.9, which is still very parallel. For
$\beta=10$, however, the situation is very different. In this case we
find at $t=0.05$ (which is the time, when the peak in the energy
evolution is most prominent) on the one hand a nearly isotropic
distribution with $a = 0.296$ for the incompressible driver. For
compressible driving on the other hand we find $a = -0.158576$, where
both of these simulations show isotropic distributions at $t =
0.3$. This shows that the compressibly driven simulations with
$\beta=10$ are the only ones, for which the tangling of the fieldlines
still changes significantly between $t=0.05$ and the saturated
state. This is also illustrated by the distribution functions shown in
Fig. \ref{FigThetaDist} for some of the simulations with $\beta =
10$. On the left we show the distribution function for very early
times, where the magnetic field is still pretty much aligned, and the
distribution for the saturated state, which is obviously quite
isotropic (here one has to take into account that an isotropic
distribution in all spatial directions yields a distribution of the
form $\sin \theta$ when projected from 3D to a
$B_{\parallel}-B_{\perp}$ distribution-function). On the right we
compare the situation for incompressible and compressible driving at
the time of the maximum. Apparently the distribution is still much
more aligned for the compressibly driven simulation.

In this context the simulations of \citet{beresnyak2009} should be mentioned: Taking into account a partially similar situation, they evaluate also multidimensional spectra, but they follow a different approach. While in this paper the global frame of reference is used (the axes are determined by the background magnetic field), they are using a local frame of reference. The discussion which of these two approaches is the best is tedious and maybe not helpful. Looking at Fig. \ref{FigThetaDist} we see that there is a clear bending of field lines, which would suggest using the local frame. But in our opinion there are two major drawbacks using the local frame: i) the definition of the local frame is not completely unique and may be governed by cutoff effects, ii) in the case of very weak turbulence local and global frame will be equivalent, while for very strong turbulence we may see differences in the local frame picture, which will in the end not affect the global picture (which is usually the one being observed).

\subsubsection{$P_{B}$}
For early times it can be seen that $P_{B}$ has higher values for
compressible than for incompressible driving (see figure
\ref{fig:subfig0.01} and table \ref{table_PBPK}). This is due to the
fact that for incompressible driving Alfv\'en waves are generated
propagating in the $x$-direction whereas for compressible
driving fast magnetosonic waves are generated propagating within the
whole perpendicular y-z-plane. So the magnetosonic waves have an
additional degree of freedom for their excitation and because of that
more fast magnetosonic waves than Alfv\'en waves are
excited for the same driving strength and driving time.\\During the
whole simulation much more turbulent energy of the velocity fields is
converted into magnetic energy for small $\beta$ than
for high $\beta$. This is due to the fact that the background magnetic field is stronger
for small plasma-$\beta$s which yields stronger magnetic fluctuations by tangling the
velocity fields.\\$P_{B}$ sometimes also shows a maximum before it
reaches the saturation range (e.g. for $\beta=0.01$ and
incompressible driving at 0.09 in figure \ref{fig:subfig0.01}). But in
this case it barely depends on the driving but on $\beta$. This is
because the rate magnetic energy is converted from the velocity fields
depends strongly on $\beta$. For $\beta=10$ this rate is anyway
so small that even a quite inefficient cascade is
adequate to remove enough energy that no maximum can build up. For
small $\beta$ the energy is converted fast enough that in principle a
maximum can build up. As for the case of incompressible driving more
energy is converted in the saturation range the
maximum is even slightly higher (see figure \ref{fig:subfig0.01} for
$\beta=0.01$ and incompressible driving at 0.09 as well as for
$\beta=0.01$ and compressible driving at 0.075 ).

\subsubsection{$P_{Shear}$ and $P_{Comp}$}
\label{energieverlaufshearcomp}
\begin {figure}
\begin{center}
\setlength{\unitlength}{0.00045\textwidth}
\subfigure[$\beta$=0.03 incompressible driving]{
    \includegraphics[width=0.45\textwidth]{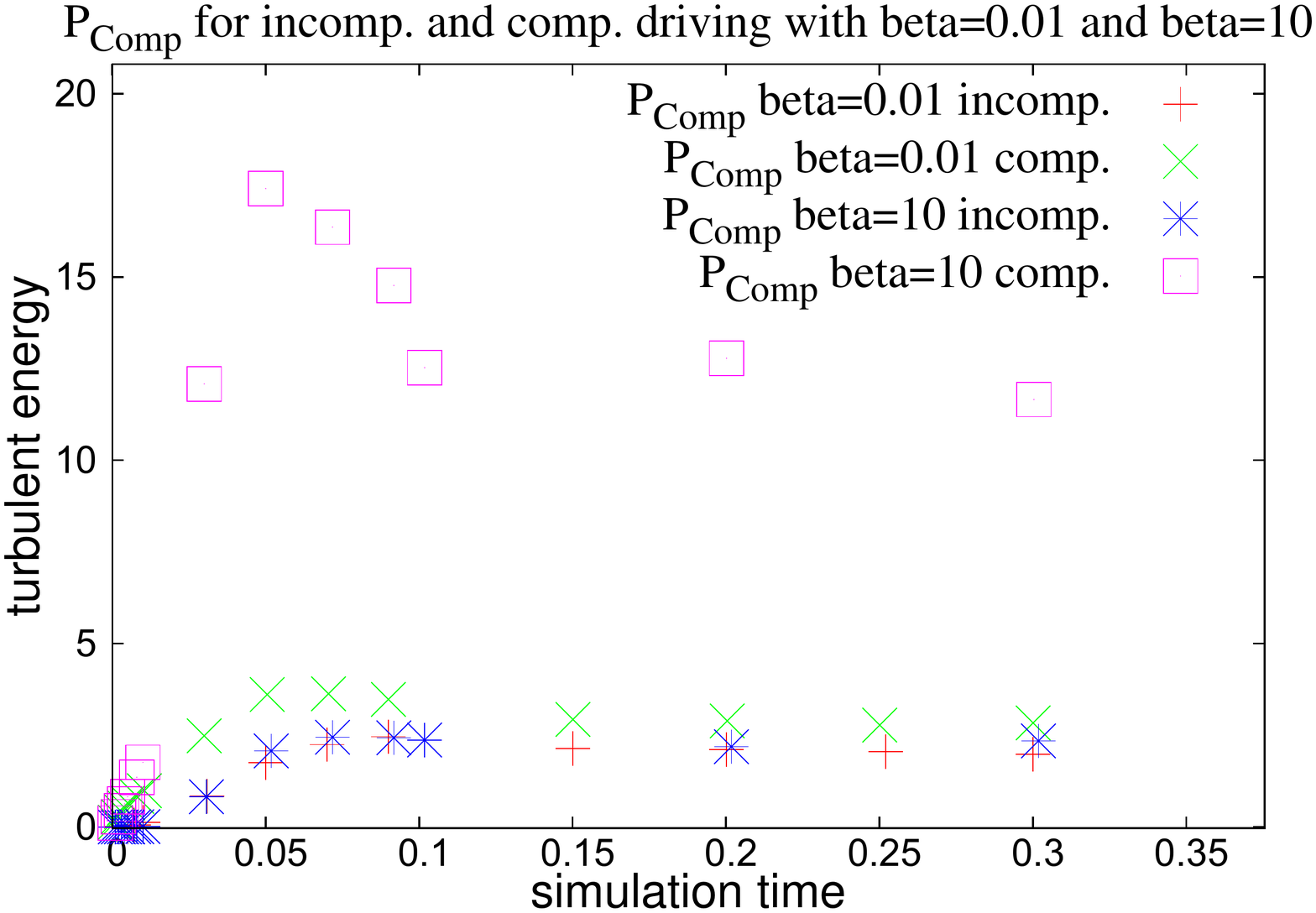}
  \label{fig:subfig0.03}
}
\subfigure[$\beta$=10 incompressible driving]{
    \includegraphics[width=0.45\textwidth]{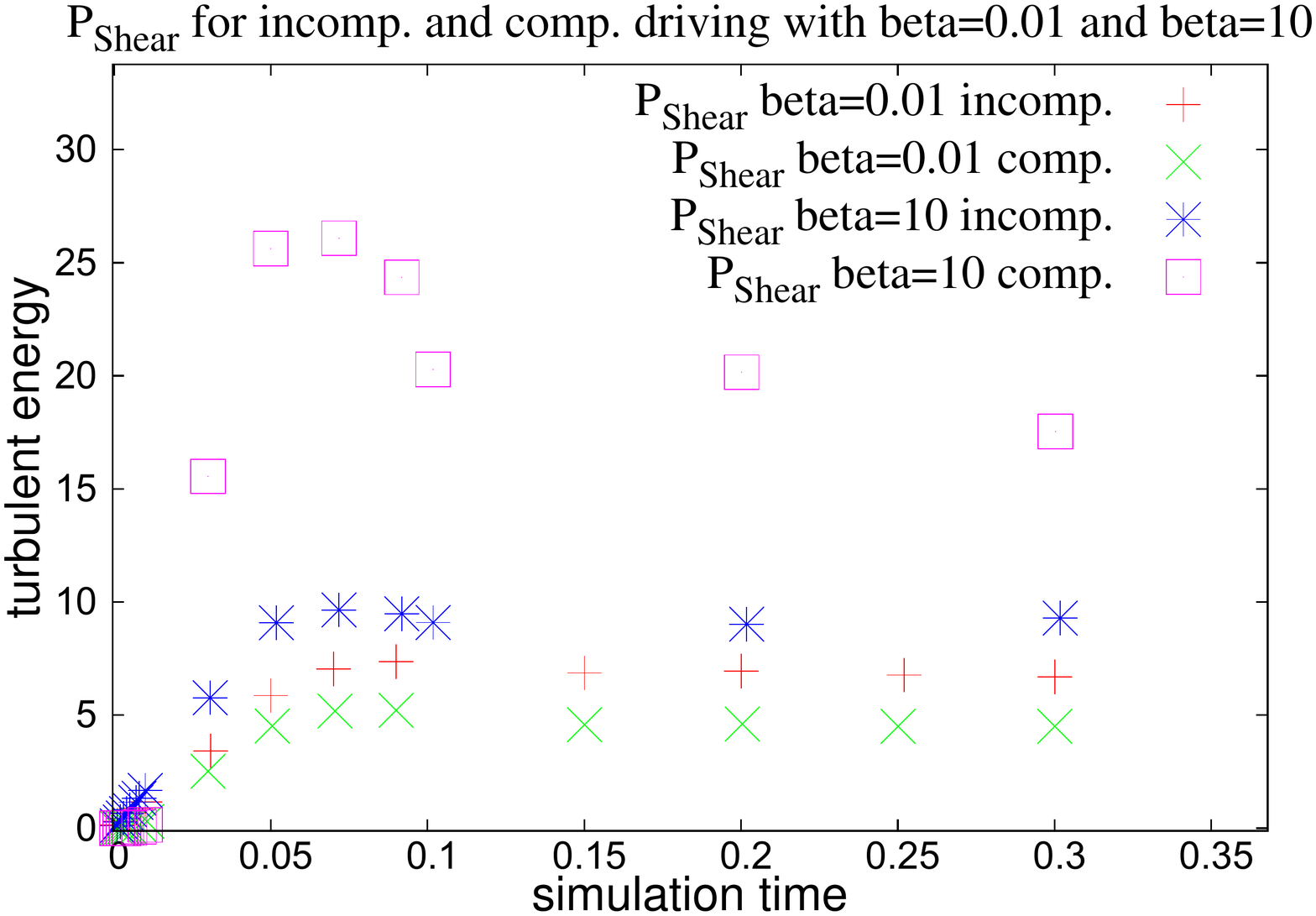}
  \label{fig:subfig0.04}
}
\end{center}
\caption{Turbulence spectra $P_{Comp}$ and $P_{Shear}$ as a function of the simulation time }
\end {figure}

\begin{figure}
\centering
\setlength{\unitlength}{0.00045\textwidth}
\subfigure[$\beta$=0.01 incompressible driving]{
  \includegraphics[width=0.45\textwidth]{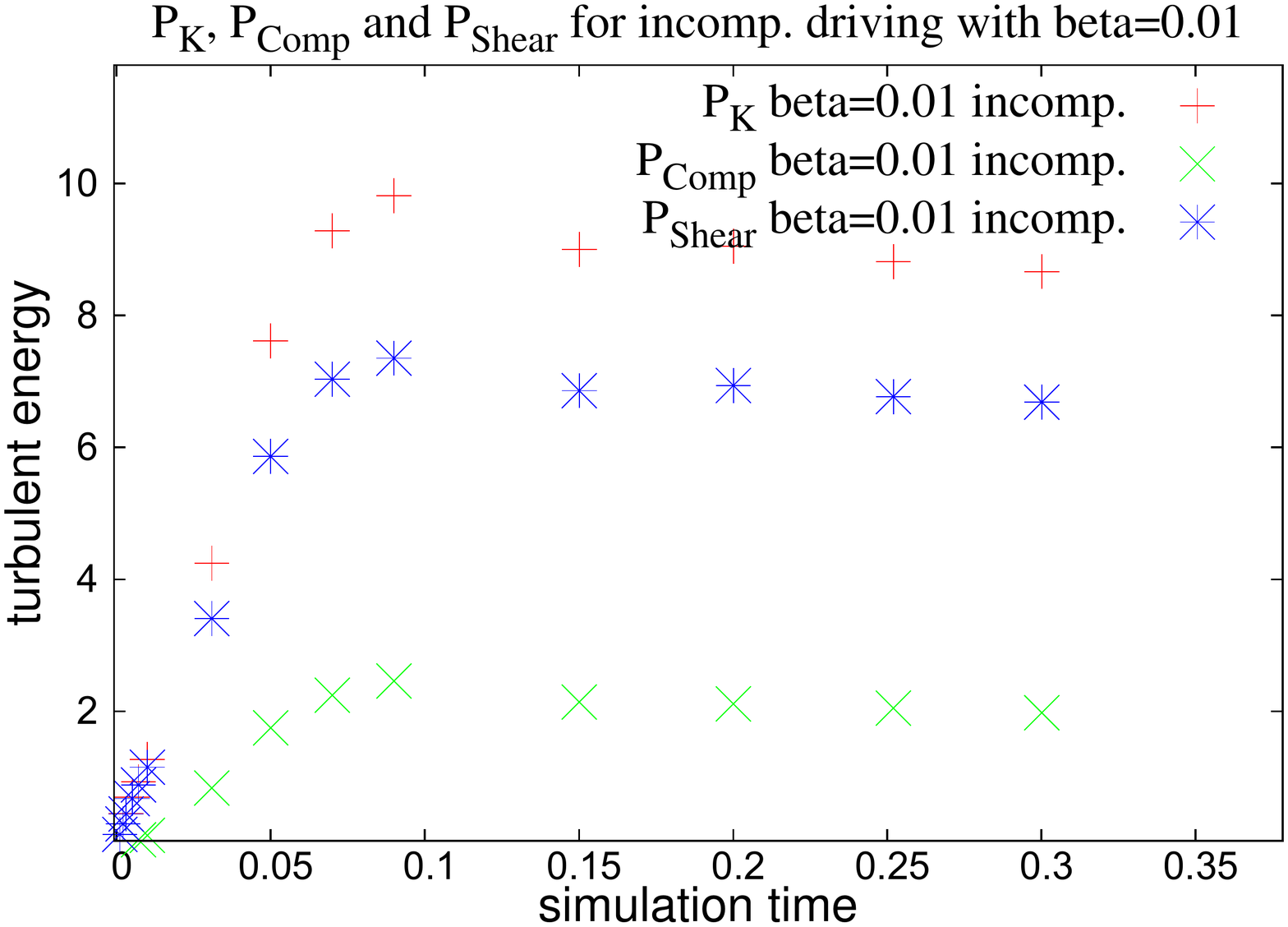}
  \label{fig:subfig3.1}
}
\subfigure[$\beta$=0.01 compressible driving]{
  \includegraphics[width=0.45\textwidth]{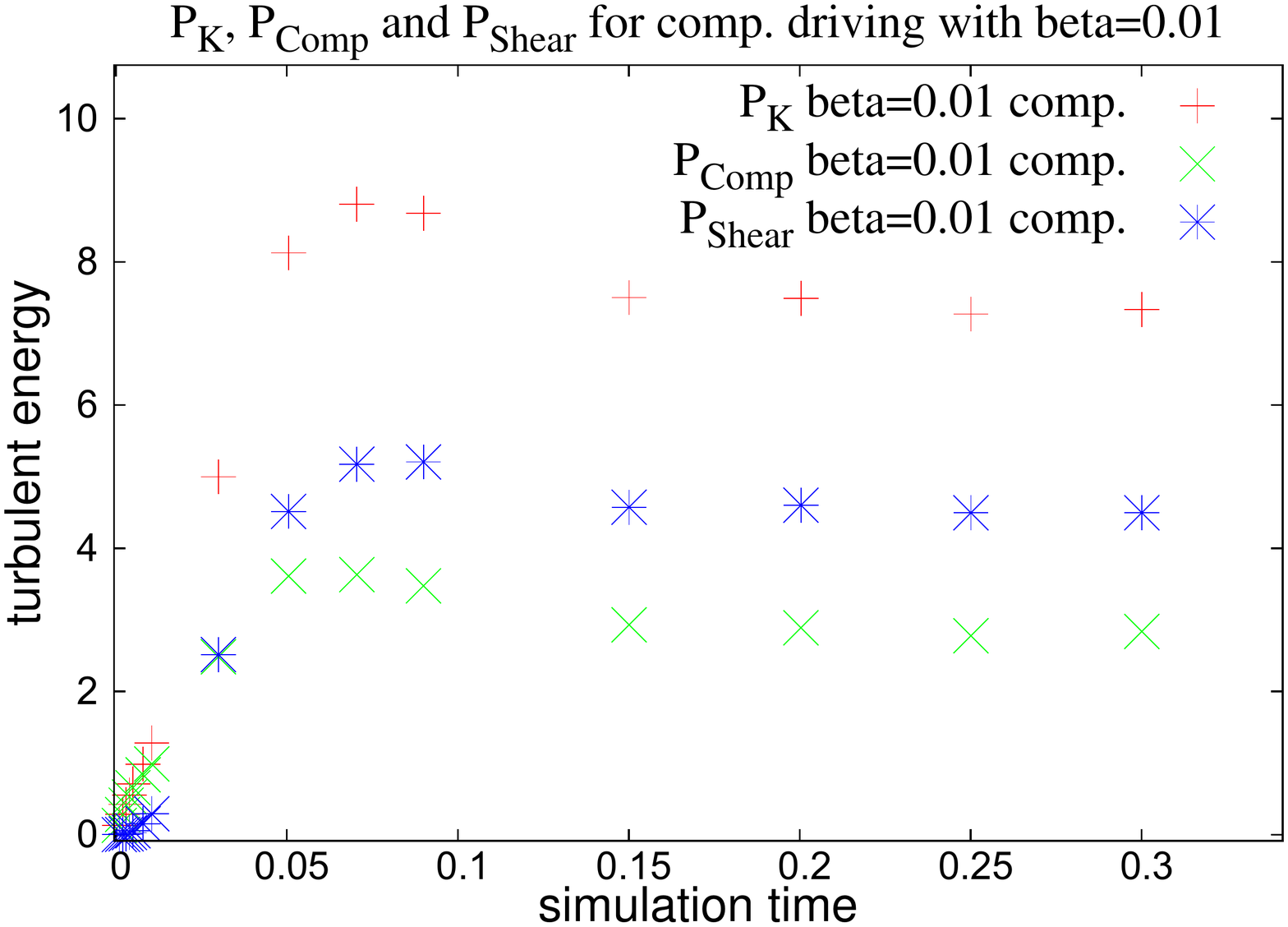}
  \label{fig:subfig3.2}
}
\subfigure[$\beta$=10 incompressible driving]{
  \includegraphics[width=0.45\textwidth]{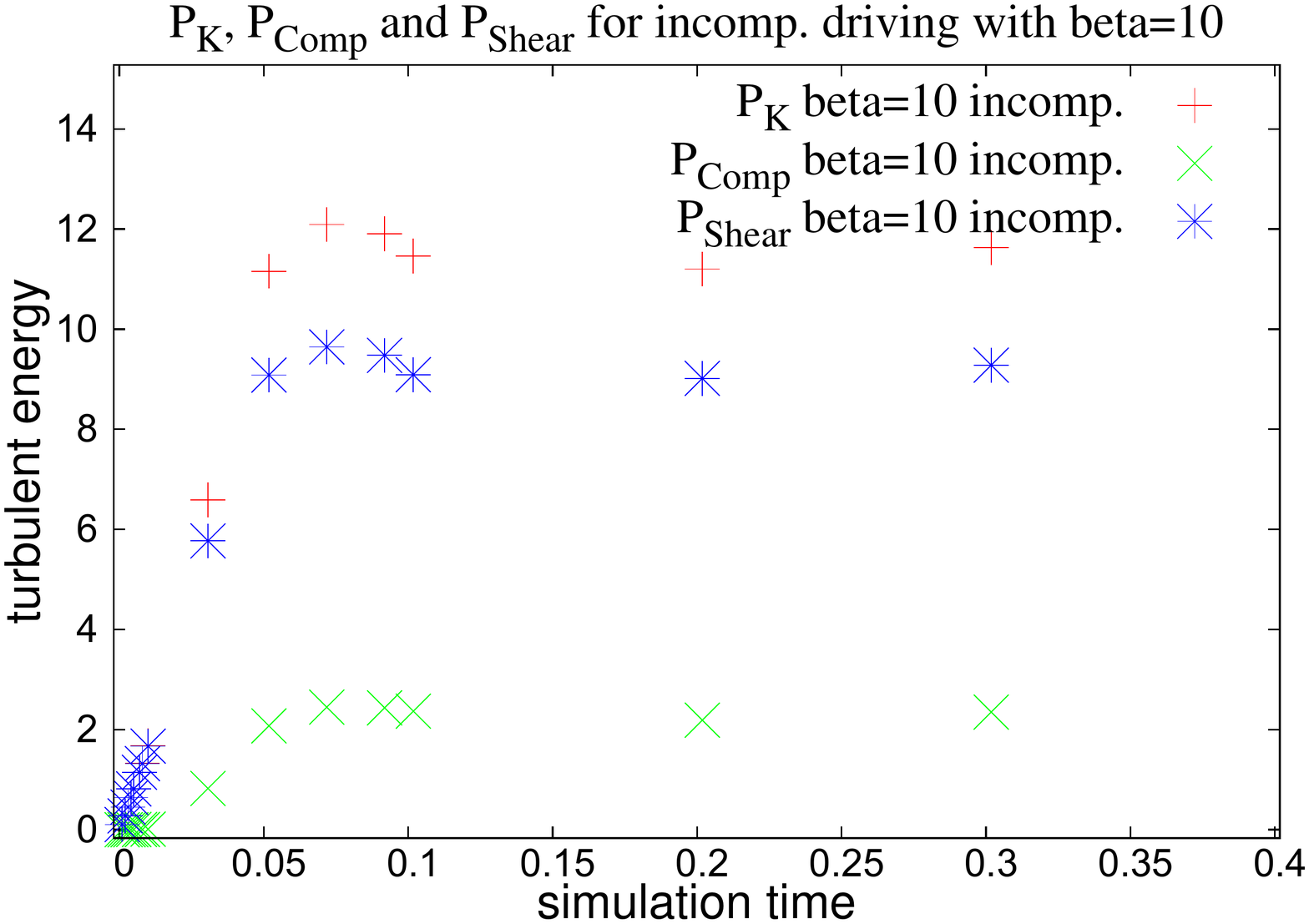}
  \label{fig:subfig3.3}
}
\subfigure[$\beta$=10 compressible driving]{
  \includegraphics[width=0.45\textwidth]{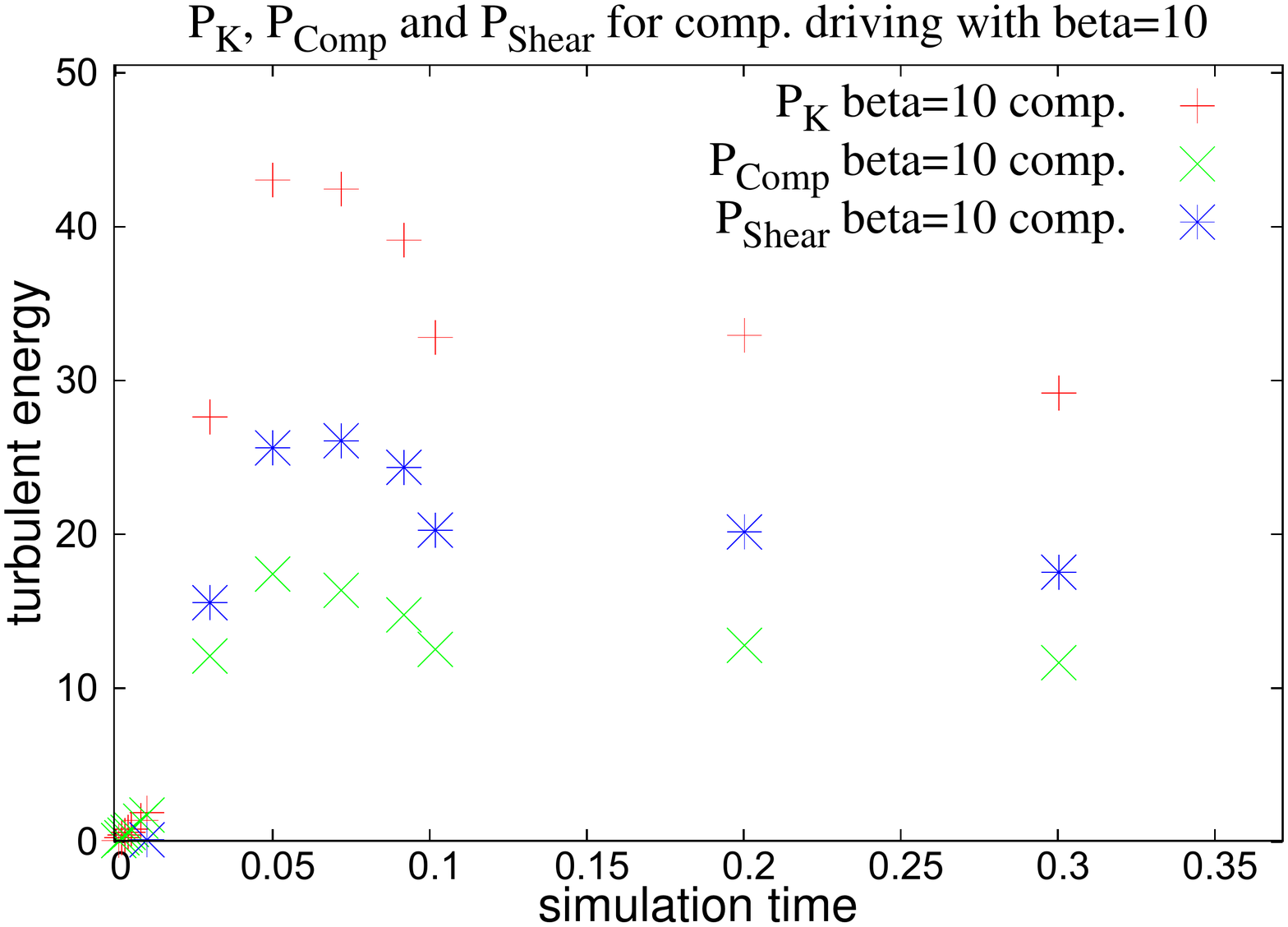}
  \label{fig:subfig3.4}
}
\caption{Total turbulent energies $P_{K}$,$P_{Comp}$ and $P_{Shear}$ as a function of the simulation time}
\end{figure}

$P_{Shear}$ and $P_{Comp}$ are plotted in \ref{fig:subfig0.03} to
\ref{fig:subfig0.04} and \ref{fig:subfig3.1} to \ref{fig:subfig3.4}
together with the total energy of the velocity fields $P_{K}$. In
table \ref{table_PShearPComp} the energies are listed for a normalised
time of $5\cdot 10^{-3}$ and for the saturated state of the spectrum.\\At the
beginning of the simulation the energy of the driving component is
higher for $\beta=10$ than for $\beta=0.01$. This is because for
bigger plasma-$\beta$s there are smaller
magnetic fields and consequently less waves are generated. So less
energy of the driving component is transformed into magnetic energy
and hence missing in the energy of the driving component. Additionally
as there are less excited modes for $\beta=10$ the non driven
component is smaller because less modes are generated that can decay
into other waves and so transform compressible modes into
incompressible modes and the other way around.\\Whereas the driving
dominates $P_{K}$ in the beginning it stands out that the
incompressible energy after some time dominates $P_{K}$ independently
of the kind of driving and the plasma-$\beta$. The dominance of the
incompressible energy might be caused by a more efficient conversion
of compressible modes into incompressible modes than the other way
around. The saturated results given in chapter \ref{chap:converge}
support this idea.

\begin {table}
  \begin{center}
    \begin{tabular}{cccccc}
 &driving&$\beta$&PB &PK&$\frac{PB}{PK}$\\\hline
time 0.05 &incomp.&0.01&9.2769$\cdot 10^{-3}$ & 0.6986&0.0133\\
time 0.05 &comp.&0.01&0.014504 &0.70617 &0.0205\\
time 0.05 &incomp.&10& 1.9613$\cdot 10^{-5}$ &0.81475  &2.407$\cdot 10^{-5}$\\
time 0.05 &comp.&10&  5.042$\cdot 10^{-5}$ & 0.828&6.089$\cdot 10^{-5}$\\\hline
saturation &incomp.&0.01&  0.177   &8.84 &0.02\\
saturation &comp.&0.01&  0.1383  &7.363 &0.0188\\
saturation &incomp.&10& 0.0675  &11.42 &5.913$\cdot 10^{-3}$\\
saturation &comp.&10& 0.0465  & 31.05&1.502 $\cdot 10^{-3}$\\
    \end{tabular}
  \end{center}
  \caption{magnetisation of the system.}
\label{table_PBPK}
\end{table}

\begin {table}
  \begin{center}
    \begin{tabular}{cccccc}
 &driving&$\beta$&PShear &PComp&$\frac{PComp}{PShear}$\\\hline
time 0.05 &incomp.&0.01  &0.676&0.022&3.25$\cdot 10^{-2}$\\
time 0.05 &comp.&0.01& 0.0521&0.654&12.55\\
time 0.05 &incomp.&10&0.814&3.19$\cdot 10^{-4}$&3.92 $\cdot 10^{-4}$\\
time 0.05 &comp.&10&6.47$\cdot 10^{-3}$&0.822&127.0 \\\hline
saturation &incomp.&0.01&6.80&2.03&0.30 \\
saturation &comp.&0.01&4.53&2.82&0.62  \\
saturation &incomp.&10&9.14&2.27&0.25\\
saturation &comp.&10&18.86&12.2&0.65 \\
    \end{tabular}
  \end{center}
  \caption{compressible and incompressible energy of the system.}
\label{table_PShearPComp}
\end{table}

\subsection{Saturated state of the spectrum}
 \label{chap:converge}

   The saturated state of the spectrum has been described in a
   number of previous papers. In the work of \citet{cho02,cho03}.
   While \citet{cho02} concentrated on the simulation of low-$\beta$
   plasmas, \citet{cho03} also considered high-$\beta$ plasmas. In
   both cases incompressible driving was used. As discussed above we
   will also use compressible driving, which may be motivated by
   supernova shock waves driving the ISM. Another point which is
   different is that the authors of the aforementioned studies
   attempted the decomposition into different wave modes. Our approach
   will focus on the decomposition in irrotational and solenoidal
   fields. This is specifically useful when one assumes that the
   turbulence is dominated by shocks.

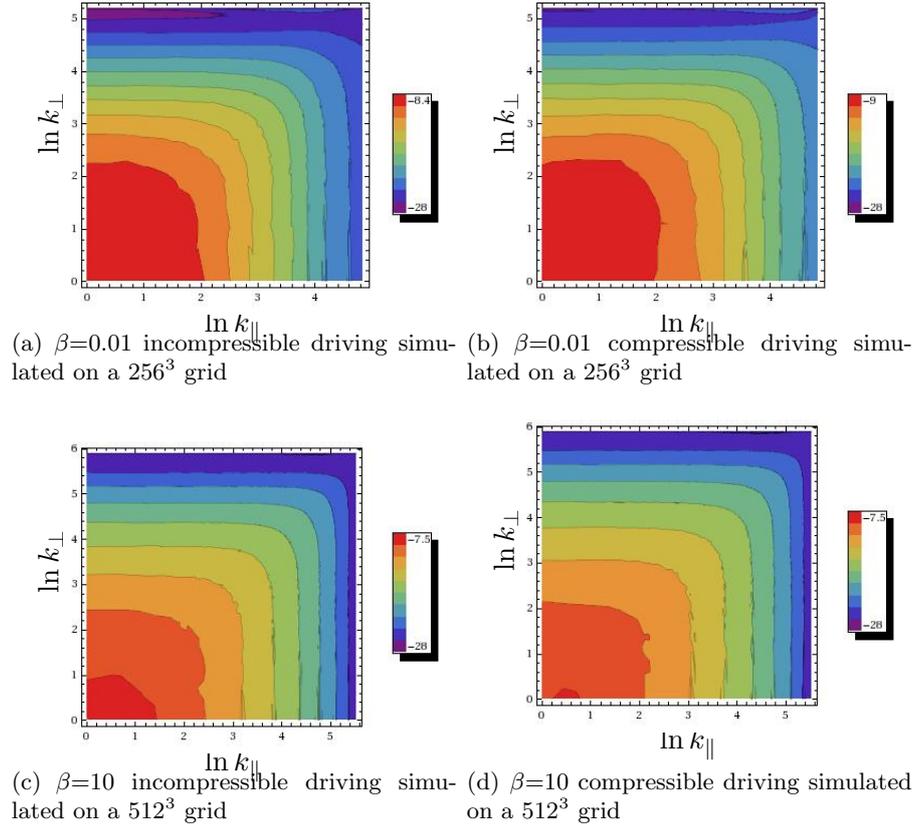
\begin{figure} \centering
\setlength{\unitlength}{0.00038\textwidth}
\subfigure[$\beta$=0.01 incompressible driving simulated on a $256^{3}$ grid]{
  \begin{picture}(1000,803)(-100,0)
    \put(-60,380){\rotatebox{90}{$\ln k_{\perp}$}}
    \put(330,-40){$\ln k_{\parallel}$}
    \includegraphics[width=900\unitlength]{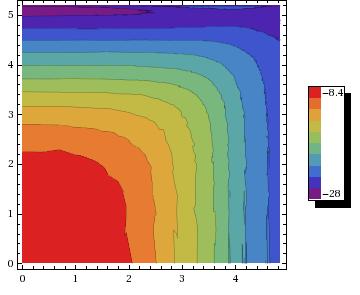}
  \end{picture}
  \label{fig:subfig9.1}
}
\subfigure[$\beta$=0.01 compressible driving simulated on a $256^{3}$ grid]{
  \begin{picture}(1000,803)(-100,0)
    \put(-60,380){\rotatebox{90}{$\ln k_{\perp}$}}
    \put(330,-40){$\ln k_{\parallel}$}
    \includegraphics[width=900\unitlength]{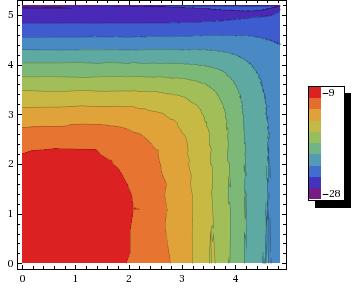}
  \end{picture}
  \label{fig:subfig9.2}
}
\subfigure[$\beta$=10 incompressible driving simulated on a $512^{3}$ grid]{
  \begin{picture}(1000,803)(-100,0)
    \put(-60,380){\rotatebox{90}{$\ln k_{\perp}$}}
    \put(330,-40){$\ln k_{\parallel}$}
    \includegraphics[width=900\unitlength]{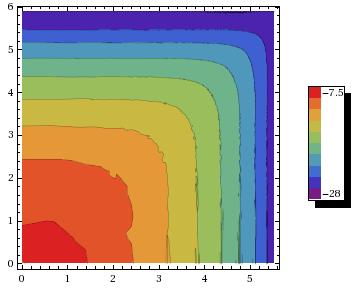}
  \end{picture}
  \label{fig:subfig9.5}
}
\subfigure[$\beta$=10 compressible driving simulated on a $512^{3}$ grid]{
  \begin{picture}(1000,803)(-100,-50)
    \put(-60,380){\rotatebox{90}{$\ln k_{\perp}$}}
    \put(330,-40){$\ln k_{\parallel}$}
   \includegraphics[width=900\unitlength]{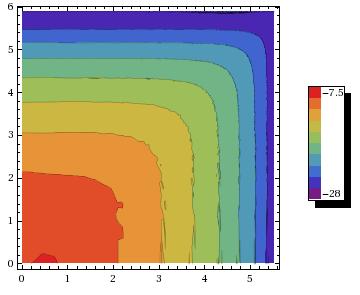}
  \end{picture}
  \label{fig:subfig9.6}
}
\caption{$P_{B}$ in the saturated state of the turbulence simulated on a $256^{3}$ and a $512^{3}$ grid.}
\end{figure}

\begin{figure}
\centering
\setlength{\unitlength}{0.00038\textwidth}
\subfigure[$\beta$=0.01 incompressible driving simulated on a $256^{3}$ grid]{
  \begin{picture}(1000,803)(-100,0)
    \put(-60,380){\rotatebox{90}{$\ln k_{\perp}$}}
    \put(330,-40){$\ln k_{\parallel}$}
    \includegraphics[width=900\unitlength]{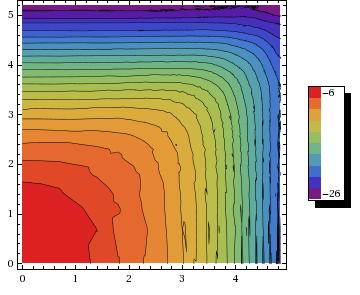}
  \end{picture}
  \label{fig:subfig7.1}
}
\subfigure[$\beta$=0.01 compressible driving simulated on a $256^{3}$ grid]{
  \begin{picture}(1000,803)(-100,0)
    \put(-60,380){\rotatebox{90}{$\ln k_{\perp}$}}
    \put(330,-40){$\ln k_{\parallel}$}
    \includegraphics[width=900\unitlength]{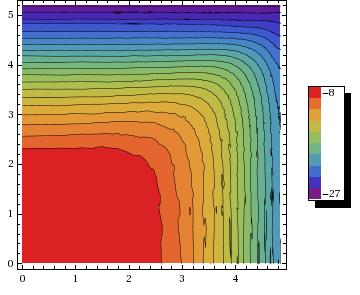}
  \end{picture}
  \label{fig:subfig7.2}
}
\subfigure[$\beta$=10 incompressible driving simulated on a $512^{3}$ grid]{
  \begin{picture}(1000,803)(-100,0)
    \put(-60,380){\rotatebox{90}{$\ln k_{\perp}$}}
    \put(330,-40){$\ln k_{\parallel}$}
    \includegraphics[width=900\unitlength]{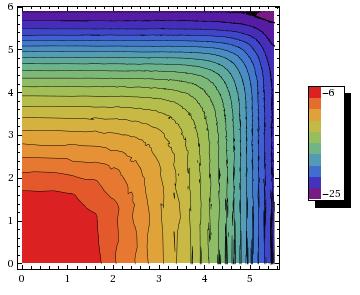}
  \end{picture}
  \label{fig:subfig7.5}
}
\subfigure[$\beta$=10 compressible driving simulated on a $512^{3}$ grid]{
  \begin{picture}(1000,803)(-100,-50)
    \put(-60,380){\rotatebox{90}{$\ln k_{\perp}$}}
    \put(330,-40){$\ln k_{\parallel}$}
    \includegraphics[width=900\unitlength]{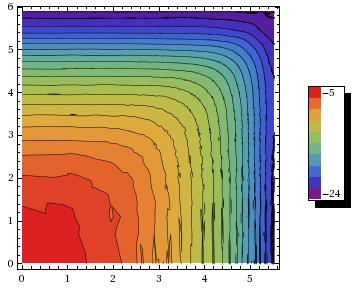}
  \end{picture}
  \label{fig:subfig7.6}
}
\caption{$P_{K}$ in the saturated state of the turbulence simulated on a $256^{3}$ and a $512^{3}$ grid.}
\end{figure}

\begin{figure}
\centering
\setlength{\unitlength}{0.00038\textwidth}
\subfigure[$\beta$=0.01 incompressible driving simulated on a $256^{3}$ grid]{
  \begin{picture}(1000,803)(-100,0)
    \put(-60,380){\rotatebox{90}{$\ln k_{\perp}$}}
    \put(330,-40){$\ln k_{\parallel}$}
    \includegraphics[width=900\unitlength]{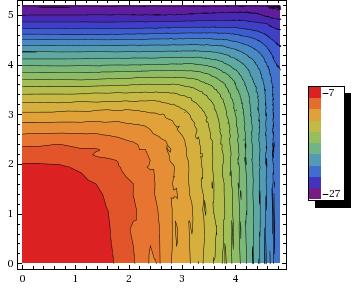}
  \end{picture}
  \label{fig:subfig10.1}
}
\subfigure[$\beta$=0.01 compressible driving simulated on a $256^{3}$ grid]{
  \begin{picture}(1000,803)(-100,0)
    \put(-60,380){\rotatebox{90}{$\ln k_{\perp}$}}
    \put(330,-40){$\ln k_{\parallel}$}
    \includegraphics[width=900\unitlength]{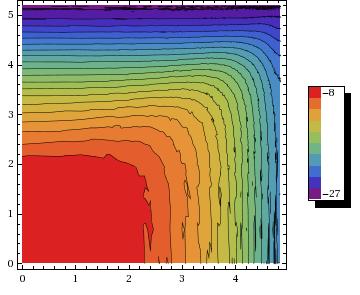}
  \end{picture}
  \label{fig:subfig10.2}
}
\subfigure[$\beta$=10 incompressible driving simulated on a $512^{3}$ grid]{
  \begin{picture}(1000,803)(-100,0)
    \put(-60,380){\rotatebox{90}{$\ln k_{\perp}$}}
    \put(330,-40){$\ln k_{\parallel}$}
    \includegraphics[width=900\unitlength]{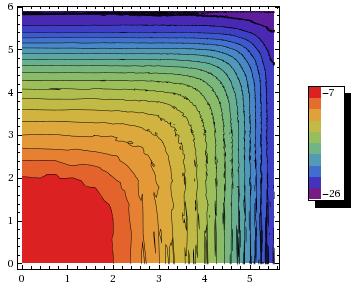}
  \end{picture}
  \label{fig:subfig10.5}
}
\subfigure[$\beta$=10 compressible driving simulated on a $512^{3}$ grid]{
  \begin{picture}(1000,803)(-100,-50)
    \put(-60,380){\rotatebox{90}{$\ln k_{\perp}$}}
    \put(330,-40){$\ln k_{\parallel}$}
    \includegraphics[width=900\unitlength]{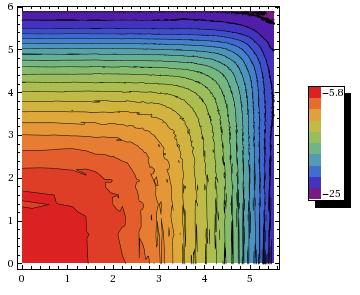}
  \end{picture}
  \label{fig:subfig10.6}
}
\caption{$P_{Shear}$ in the saturated state of the turbulence simulated on a $256^{3}$ and a $512^{3}$ grid.}
\end{figure}

\begin{figure}
\centering
\setlength{\unitlength}{0.00038\textwidth}
\subfigure[$\beta$=0.01 incompressible driving simulated on a $256^{3}$ grid]{
  \begin{picture}(1000,803)(-100,-50)
    \put(-60,380){\rotatebox{90}{$\ln k_{\perp}$}}
    \put(330,-40){$\ln k_{\parallel}$}
    \includegraphics[width=900\unitlength]{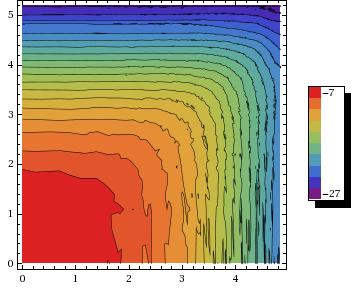}
  \end{picture}
  \label{fig:subfig8.1}
}
\subfigure[$\beta$=0.01 compressible driving simulated on a $256^{3}$ grid]{
  \begin{picture}(1000,803)(-100,-50)
    \put(-60,380){\rotatebox{90}{$\ln k_{\perp}$}}
    \put(330,-40){$\ln k_{\parallel}$}
    \includegraphics[width=900\unitlength]{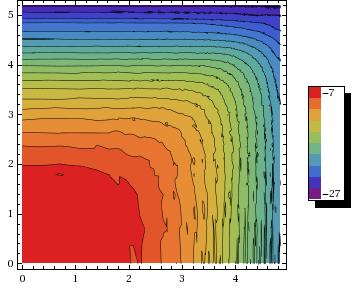}
  \end{picture}
  \label{fig:subfig8.2}
}
\subfigure[$\beta$=10 incompressible driving simulated on a $512^{3}$ grid]{
  \begin{picture}(1000,803)(-100,-50)
    \put(-60,380){\rotatebox{90}{$\ln k_{\perp}$}}
    \put(330,-40){$\ln k_{\parallel}$}
    \includegraphics[width=900\unitlength]{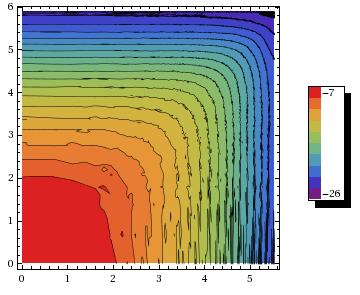}
  \end{picture}
  \label{fig:subfig8.5}
}
\subfigure[$\beta$=10 compressible driving simulated on a $512^{3}$ grid]{
  \begin{picture}(1000,803)(-100,-50)
    \put(-60,380){\rotatebox{90}{$\ln k_{\perp}$}}
    \put(330,-40){$\ln k_{\parallel}$}
    \includegraphics[width=900\unitlength]{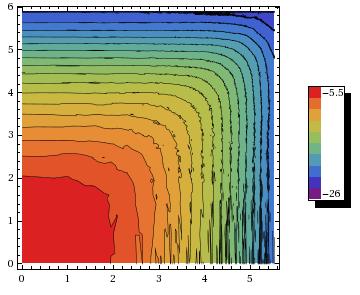}
  \end{picture}
  \label{fig:subfig8.6}
}
\caption{$P_{Comp}$ in the saturated state of the turbulence simulated on a $256^{3}$ and a $512^{3}$ grid.}
\end{figure}

Here we present the two dimensional spectra for the saturated state of
the turbulence. For $\beta=0.01$ and compressible driving $P_{K}$,
$P_{Shear}$ and $P_{Comp}$ have a clearly parallel preferred direction
(see figure \ref{fig:subfig7.2},\ref{fig:subfig10.2} and
\ref{fig:subfig8.2}).  In the saturated state we are concerned with
fully nonlinear interactions, which form an anisotropic cascade, as we
see from the results.  For incompressible driving we find that $P_{K}$
and $P_{Shear}$ develop a perpendicular preferred direction (see
figure \ref{fig:subfig7.1} and \ref{fig:subfig10.1}), as it has been
predicted by \citet{shebalin83}.  The spectrum of $P_{Comp}$ however is
quite isotropic (see figure \ref{fig:subfig8.1}). As it has been
already mentioned in chapter \ref{energieverlaufshearcomp}, we believe
that compressible modes may easily be converted into incompressible
modes, but not the other way around. This is confirmed by the two
dimensional spectra, because for compressible driving $P_{Shear}$ has
the same anisotropy as $P_{Comp}$. Actually the incompressible modes
would build up a perpendicular cascade as can be seen by the results
in \ref{fig:subfig7.1} and \ref{fig:subfig10.1}, as they are however
directly generated by the compressible modes, the cascade is forced to
build up the same preferred direction. Contrariwise in the case of
incompressible driving $P_{Comp}$ has not the same preferred direction
as $P_{Shear}$. So the transformation of compressible modes from
incompressible modes is not that efficient. This is why a
superposition of compressible and incompressible modes and so of a
parallel and perpendicular cascade can be seen. The result is an
isotropic spectrum.
\\
For $\beta=0.01$ it can be seen that for incompressible driving
$P_{B}$ has the same perpendicular preferred direction that we also
find for $P_{K}$, $P_{Shear}$ (see figure \ref{fig:subfig9.1}) whereas
for compressible driving there is no preferred direction (see figure
\ref{fig:subfig9.2}). In figure \ref{fig:subfig3.1} it can be seen
that for incompressible driving much more incompressible modes than
compressible modes are in the system. This is why we would also expect
a perpendicular cascade to build up. For the case of compressible
driving (see figure \ref{fig:subfig3.2}) we have only slightly more
incompressible than compressible modes in the system. In this case the
magnetic energy will cascade in parallel and perpendicular
direction. One requirement is that magnetic and kinetic
  energy may cascade differently. The spectra suggest that this is indeed what happens in the simulations,  but an explanation is hard.

For $\beta=10$ the two dimensional spectra are quite isotropic as we
would predict it for a this widely hydrodynamical case (see figure
\ref{fig:subfig9.5}, \ref{fig:subfig9.6}, \ref{fig:subfig7.5},
\ref{fig:subfig7.6}, \ref{fig:subfig10.5}, \ref{fig:subfig10.6},
\ref{fig:subfig8.5} and \ref{fig:subfig8.6}).

Similar investigations have been done by \citet{2003ApJ...590..858V} for the case of
incompressible driving in the saturated state of the spectrum. He
also found a perpendicular preference of the fluctuations for small
values of the plasma-$\beta$, which disappears for higher $\beta$.

\section{Discussion}

In this paper we presented results from simulations of compressible
MHD turbulence for low- and high-beta plasmas with compressible and
incompressible driving. Our special focus was on the evolution of the
turbulence. We found an obvious influence of the plasma-$\beta$ and
the type of driving on the anisotropy and the temporal energy
evolution. Especially the influence of the driver has not yet
  been discussed for MHD plasmas so far.\\ This may have influence on
a number of astrophysical scenarios. When assuming supernovae as
energy input, it is evident that e.g.~ the pile-up of
fluctuation-energy due to slow tangling of magnetic field lines may be
observed in the early stages as a lower temperature of the plasma. So
by observing temperature evolution in turbulent driving it should be
possible to determine which process is really driving the turbulence:
Incompressible or compressible fluctuations.  \\ Even more striking is
the influence on particle acceleration in evolving turbulence. Here
also early supernova remnants are the object of interest. The first
thing to observe here, is that a weak cascade in compressible driving
may prevent low energetic particles to be accelerated due to missing
energy at their respective resonant high wavenumbers.\\ It is not
completely clear if the anisotropy arising in first place will have an
observable effect, but it is very well possible that the anisotropy in
the magnetic fluctuations may change the transport and consecutively
the acceleration and escape of cosmic ray from turbulent
regions.\\

\section{Summary}

This paper gave summary of the evolution of turbulence in a driven MHD plasma. The evolution is described especially under the influence of different driving mechanisms (compressible and incompressible). First principles are able to explain the spectra for the very early development of turbulence. The anisotropy of wave generation is able to explain the specific features visible. For the further development of turbulence the plasma beta plays an essential role in the saturation of the spectra. We have shown the distribution of energy between magnetic and kinetic energy depending on the plasma beta. The results shown here are in agreement with earlier research done for hydrodynamics and goes into more detail regarding the specific anisotropy introduced with the background field.

\bibliographystyle{jpp}

\section*{Acknowledgements}
      MW acknowledges support by Graduiertenkolleg 1147 and FS acknowledges support by
      \emph{Deut\-sche For\-schungs\-ge\-mein\-schaft, DFG\/} project
      number Sp~1124/3.\\
      We would like to thank the anonymous referee for his detailed comments, which
      helped improving this paper.

\begin{appendix}
  \section{Measure of the anisotropy}
\label{app:aniso}  
  Anistropy is one of the main features discussed in the context of this paper. The qualitative behaviour of the anisotropy can be seen in the two-dimension plots of the spectrum. A quantitative description is far more difficult since the actual shape of the two-dimensional spectra may differ strongly.
  
  To give a rough quantitative comparison we are trying to boil down the plot to single numbers. We are using the contour lines to find a $k_\perp^\text{aniso}(k_\parallel)$. By following a contour line starting at a given value of $k_\parallel$ from the $k_\parallel$ axis to the $k_\perp$ axis we find this parameter.
  
\begin{figure}
    \includegraphics[width=.9\textwidth]{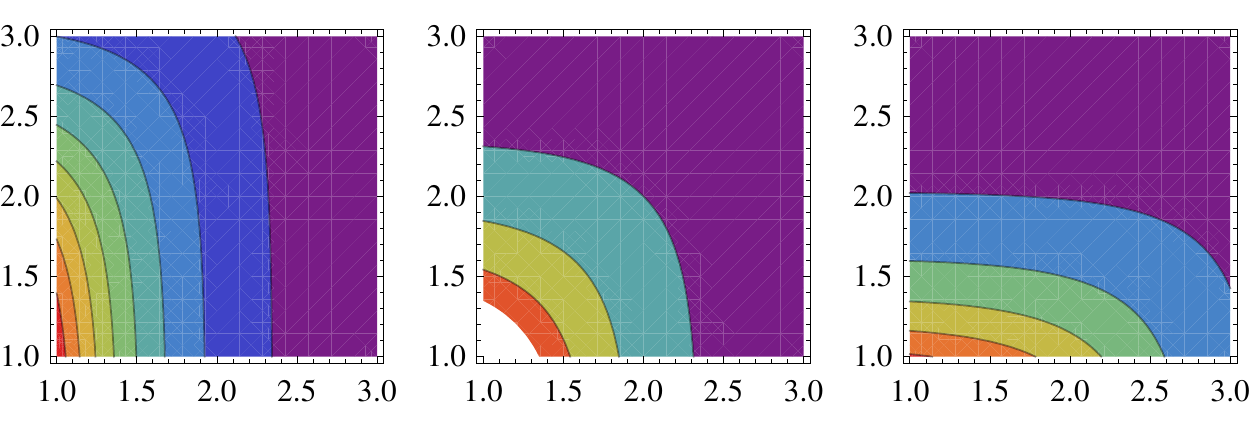}
    \caption{Two-dimensional spectra for the sample spectrum Eq. \ref{eq:sample}. Shown are plots for the values of $\Lambda=0.1, 1, 10$.}
    \label{fig:sample}
  \end{figure}

  We want to illustrate the determination with a sample anisotropic function
  \begin{equation}
    f(x,y) = \frac{1}{(x^2 + \Lambda y^2)^{\frac{5}{3}{1}{2}}}
    \label{eq:sample}
  \end{equation}
  The contour plots for different parameters $\Lambda$ are shown in Fig. \ref{fig:sample}. For the simple case of $\Lambda=1$ we find for example $k_\perp^\text{aniso}(2.3) = 2.3$. The ratio of $k_\parallel$ and $k_\perp$ is here 1, since the function is anisotropic. For $\Lambda = 0.1$ we find $k_\perp^\text{aniso}(1.7) = 2.7$ and for $\Lambda = 10$ we find $k_\perp^\text{aniso}(2.6) = 1.6$.
  
This number gives a first quantitative hint on the anisotropy of the problem, but it is limited to axis aligned asymmetry. For feature propagating at $45$ degree to the magnetic field it does not show any features.   
  
\end{appendix}

\bibliography{apj-jour,paperbib}

\end{document}